%% file: Void_SALT_p2.tex
\documentclass[useAMS,usenatbib]{mn2e}
\usepackage{graphicx}
\usepackage{amssymb}
\usepackage{rotating}
\usepackage{float}

\newcommand{\apj}{ApJ}

\newcommand{\mnras}{MNRAS}

\newcommand{\MC}{\multicolumn}
\newcommand{\kms}{km~s$^{-1}$}

\newcommand{\HI}{H{\sc i}}
\newcommand{\HII}{H{\sc ii}}
\newcommand{\sunn}{$_{\odot}$}
\newcommand{\p}{$\pm$}

\newcommand{\dg}{$\dagger$}
\newcounter{qub}
\newcommand{\qq}{\addtocounter{qub}{1}\arabic{qub}}

\title[Dwarfs in nearby voids: SALT spectroscopy]
{Dwarfs in nearby voids: results of SALT spectroscopy}

\author[S.A.~Pustilnik, A.Y.~Kniazev, A.L.~Tepliakova, Y.A.~Perepelitsyna,  E.S.~Egorova]
{S.A.~Pustilnik,$^{1}$\thanks{E-mail: sap@sao.ru (SAP)} A.Y.~Kniazev,$^{2,3,4}$ A.L.~Tepliakova,$^{1}$
  Y.A.~Perepelitsyna,$^{1}$ E.S.~Egorova$^{4,5}$  \\
$^1$ Special Astrophysical Observatory of RAS, Nizhnij Arkhyz,Karachai-Circassia 369167, Russia \\
$^2$ South African Astronomical Observatory, PO Box 9, 7935 Observatory,  Cape Town, South Africa \\
$^3$ Southern African Large Telescope Foundation, PO Box 9, 7935 Observatory, Cape Town, South Africa \\
$^4$ Sternberg Astronomical Institute, Lomonosov Moscow State University, Universitetskij Pr. 13,  Moscow 119992, Russia \\
$^5$ Astronomisches Rechen-Institut, Zentrum f\"{u}r Astronomie der Universit\"{a}t Heidelberg, M\"{o}nchhofstra\ss e 12-14, 69120 Heidelberg, Germany
}

\begin{document}

\label{firstpage}

\date{Accepted December ??, 2023, Received August 20, 2023}

\pagerange{\pageref{firstpage}--\pageref{lastpage}} \pubyear{2023}

\maketitle

\begin{abstract} 
In the framework of the ongoing project, aimed at the systematical studying galaxies in nearby voids,
we conducted spectroscopy with the Southern African Large Telescope (SALT)
of 62 objects from the Nearby Void Galaxy (NVG) sample. They include 8 remaining objects of
the 60 preselected candidates to eXtremely Metal-Poor (XMP) dwarfs, two known void XMP dwarfs
and 52 void dwarfs residing within the Local Volume. For 47 galaxies residing in the nearby voids,
we obtained spectra of the diverse quality. For 42 of them, we detected the
Hydrogen and Oxygen lines that allowed us to get estimates of O/H in the
observed \HII\ regions.
For 12 of the 42 objects, we detected the faint line [O{\sc iii}]$\lambda$4363,
that allowed us to directly derive the electron temperature T$_{\rm e}$ and obtain their
gas O/H by the direct method.
14 objects with the undetected [O{\sc iii}]$\lambda$4363 line fall to the lowest metallicities
range (12+log(O/H) $\lesssim$ 7.5~dex). For them, we use a carefully checked
new empirical 'Strong line' method of Izotov et al.
For 14 other objects with only strong lines detected and with 12+log(O/H) of $\sim$7.5--8.0~dex,
we used the modified version of 'semi-empirical' method of Izotov and Thuan. It accounts for effect of
the excitation parameter O$_{\mathrm 32}$ on T$_{\rm e}$.
16 new galaxies are found with parameter 12+log(O/H) $\lesssim$ 7.39~dex. Of them, four have
12+log(O/H) = 7.07 -- 7.20~dex. Of the 60 observed NVG objects, 15 have mistaken radial velocities in HyperLEDA.
They do not reside in the nearby voids.
\end{abstract}


\section[]{Introduction}
\label{sec:intro}
\setcounter{figure}{0}

Voids represent the most rarefied and the largest volume elements of the
Large-Scale Structure of the Universe.
Despite galaxies residing in voids comprise a relatively small fraction
of the whole realm of galaxies \citep{Weygaert16},
they are the important 'bricks' of our picture of the general galaxy evolution.
Thus,  \citet{Aragon13} predict the unusual properties of void structures and their
galaxies due to the 'time machine and cosmic microscope' effects of the void environment.

A more detailed modelling of the void galaxy evolution is necessary to
compare their predicted properties with
the observed ones. In particular, one needs the high resolution simulations
in order to address the small
dwarfs with the baryonic mass of 10$^6$ -- some 10$^8$ M\sunn. Namely for
this mass range, the observations (see below)
demonstrate significant differences with the properties of galaxies in denser environments.

  The effect of environment on evolution is expected to be stronger for the lower mass
   objects. Since all known wide-area spectral sky surveys have the similar limiting apparent
   magnitude of B (or $r$) of $\sim$18.5, to have a representative galaxy sample to, e.g.,
   M$_{\rm B}$ = --10 to --12, one should explore the volume with the distances of R $\lesssim$ 20--25 Mpc.
   This choice differs substantially from many studies of void galaxy samples in more distant voids
   ($\sim$80--200~Mpc), which probe properties of only the upper part of the galaxy luminosity function
   \citep[e.g.,][and references therein]{Rojas05,Patiri06,Kreckel12}.

An alternative approach related to the study of galaxies in the nearby voids, was suggested and partly
realised by our group. The first step in this direction was the formation and studying of the
galaxy sample in the nearby void Lynx-Cancer  \citep[][and references therein]{PaperI, PaperVII}.
  In particular, it was found that void galaxies as a whole are less evolved in average
  as compared to the reference sample of the Local Volume  (LV) late-type galaxies of \citet{Berg12}.
  They have elevated, in average, ratio of gas mass to luminosity (M(\HI)/L$_{\rm B}$) \citep{PaperVI} and
  the reduced gas metallicity \citep{PaperVII, XMP.BTA}.

A surprising result of the above unbiased study of the whole Lynx-Cancer void sample was
the discovery of several low-luminosity dwarfs with the eXtremely Metal-Poor (XMP) \HII\ regions
and the extremely-high gas mass fraction, f$_{\rm g}$ = \mbox{M$_{\rm gas}$/M$_{\rm baryon}$},
up to 0.97--0.99 \citep{J0926,PaperIII,Triplet,U3672}. Besides, an independently discovered XMP dwarf
AGC198691  \citep{Hirschauer16,Aver2022} appeared to reside in the same void. We adopt here
the definition of XMP galaxies as
objects with 12+log(O/H) $\lesssim$ 7.21~dex, or with Z(gas) $\lesssim$ Z\sunn/30\footnote{
The solar value of 12+log(O/H) is adopted to be 8.69~dex after \citet{Asplund09}.}.
This metallicity is close to that of the prototype of such unusual galaxies,
the famous blue compact dwarf  IZw18 \citep{Searle72}.

In addition, these unusual dwarfs reveal blue colours of the outer parts of a galaxy
\citep[e.g.][]{PaperIV,U3672},
where, in the framework of the common paradigm, the most aged stellar population resides.
The colours, being compared to the evolutionary tracks from PEGASE2 package \citep{PEGASE2},
correspond to the ages of the related stellar population of about one to several Gyr.
That hints, along with the extremely high gas mass fraction, to the early stages of galaxy evolution.

These properties, in turn, may connect the unusual void dwarfs with the predicted so-called
Very Young Galaxies (VYG) \citep{Tweed18}, defined as objects, formed more than a half of stellar mass
within the last 1~Gyr. While the other their properties are not specified in simulations, the first
attempts to find such a rare population are presented by \citet{Mamon20}. Meanwhile, the comprehensive
study of a void XMP dwarf IZw18 and its companion IZw18C gives a clear indication on their
VYG nature \citep{PO12}.

The seach for and the study of XMP dwarf galaxies was in focus of many groups since the
discovery
of the record low metallicity of IZw18. We already discussed recently the continuing
efforts and findings of such unusual galaxies in the previous papers of this series. Therefore we
refer a reader to these texts. The only remarks we would add to this issue, are the recent results of
the search for new XMP dwarfs within the nearby voids \citep{XMP.SALT, XMP.BTA}. They appeared very
promising, since the number of known such objects was doubled thanks to our conducted program
during several last years.

However, one should note that the prominent XMP dwarfs in the local Universe
are occasionally found also in a less rarefied environments. The known examples include
'Little Cub' galaxy \citep{LittleCub} and 'Peekaboo' galaxy \citep{Peekaboo} with
12+log(O/H) = 7.11 and 6.98~dex, respectively.
This is not clear yet, whether their evolutionary paths differ from those for the
similar objects in voids.

\input{tab_jour1.tex}

The voids themselves are not simple and homogeneous. They have the fine sub-structure resembling
the Large-Scale Structure of the Universe \citep[e.g.][]{Gottlober03, Aragon13}. This sub-structure
contains nodes, filaments and sub-voids.
Therefore, one expects that properties of void galaxies can vary, in particular due to the
different effects of their in-void {\it local} environments.

The first results, briefly summarised above, appeal to a much more massive studying of void dwarf galaxies.
To address the diversity of void galaxies and to have an opportunity to compare cosmological simulations with
the real void objects, one needs properties for a sample of many hundreds to thousands galaxies within
the limited volume, unbiased as much as possible.

The natural questions related to such studies are the following:
a) how much properties of void galaxies can differ from those in a more typical environment?
b) how often properties of void galaxies {\it significantly} differ from those of a more common population?
c) what empirical relations exist between the properties of void galaxies and other factors.
The answers to these questions can help us to understand possible differences in formation and evolution
scenarios of void galaxies from those in a more common environment.

As a development of the observational study of dwarf void galaxies, in order to increase
substantially the statistics of such objects with known evolutionary parameters, Z(gas) and
gas mass fraction, we formed a sample of 25 voids in the volume with R $<$ 25 Mpc from
the Local Group. We separate 1354 galaxies within this volume ($\sim$20 percent of the total
number known in this volume) which fall within these nearby voids.
We refer to this sample as the 'Nearby Void Galaxy' (NVG) sample \citep[][hereafter PTM19]{PTM19}.
The sample galaxies were collected mostly from the HyperLEDA data base \citep{LEDA}.

The main parameters of the parent NVG sample are as follows. The range of the absolute
blue magnitudes is of M$_{\rm B}$ = --7.5 to --20.5~mag, with the median value of about --15.2~mag.
The great majority of the NVG sample are the late-type galaxies (late spirals and dwarf
irregulars), with the major fraction of the low surface brightness (LSB) galaxies. Similar to
the wide range of luminosities, the distribution on the hydrogen mass in void galaxies is also very broad,
with the full range of M(\HI) of 10$^{5.5}$ -- 10$^{9.7}$~M\sunn, with the median M(\HI) $\sim$10$^{8.5}$~M\sunn.
The parameter of gas richness, the ratio M(\HI)/L$_{\rm B}$ of about a thousand NVG sample galaxies with
known \HI\ data, spreads from 0.03 to 26, with the median of $\sim$1.

In the procedure of separation of void galaxies, all objects, appeared in the preselected empty
spheres, were assigned to the void sample. They were further divided to the 'inner' void subsample
($\sim$80~\%),  and the 'outer' subsample. The 'inner' 1088 galaxies reside in the 'inner' parts of
voids. Their distances to the nearest luminous neighbours, delineating voids, D$_{\rm NN} \geq$ 2.0~Mpc,
with the median value of $\sim$3.4~Mpc.

The 'outer' 242 galaxies have D$_{\rm NN} < $ 2.0~Mpc, with
the median of 1.65~Mpc. The separation of the void subsamples at D$_{\rm NN} = $ 2.0~Mpc was chosen
to be consistent with the definition of galaxies residing in the nearby Lynx-Cancer void from
our previous study \citep{PaperI, PaperVII}. In that work, the border of D$_{\rm NN} > $ 2.0~Mpc was
intentively chosen so that to study the well defined void objects and to exclude possible interlopers.
The study of both, the 'inner' and the 'outer' subsamples, will allow one to understand how many
of the 'outer' void galaxies represent on their properties an 'intermediate' kind of the galaxy population
in the context of transition
between the lowest density regions and the walls, with a roughly mean matter density.

We then began two projects based on this sample. One of them is the search for new XMP void dwarfs.
We selected 60 candidate XMP dwarfs from the whole NVG sample based on their known properties
in the open data bases and in the literature \citep{PEPK19}. The results of
spectroscopy of $\sim$3/4 of them are presented in papers \citet{XMP.SALT, XMP.BTA}.

Another project is devoted to the study of the subsample of the NVG objects which fall
within the LV. This consists of $\sim$260 objects from the NVG at the distances closer than 11~Mpc.
First results on this part were presented in  \citet{KK242, MUTO22}.
This is an analogue of our unbiased study of the galaxy sample in the nearby void Lynx-Cancer.
The northern part of void galaxies in this project is studied at the SAO 6-m telescope BTA.
The detailed description
of the LV void sample, with some intermediate results, will appear in a paper in preparation.

 Here we present the results of
the Southern African Large Telescope \citep[SALT;][]{Buck06,Dono06}  spectroscopy
from the both projects.
The first part includes  eight of the remaining 14 dwarfs selected in \citet{PEPK19} as the void XMP
candidates. The second part includes 49 galaxies selected as residing in voids within the LV, mainly
the least luminous dwarfs. Also, a known XMP dwarf J0015+0104 from the Eridanus void
\citep{Guseva09, Pustilnik13} was observed to improve the accuracy of its O/H.

The content of the paper is arranged as follows:
the description of the SALT spectral observations and data processing
is presented in Sec.~\ref{sec:observing}.
In Sec.~\ref{sec:OH} we give the description of emission line measurements and methods
used for O/H determination.
The estimates of O/H for the observed galaxies are shown in Sec.~\ref{sec:results}.
In Sec.~\ref{sec:discussion} we discuss the obtained results along with other
available information.
In Sec.~\ref{sec:conclusions} we  summarise and  conclude.
In the on-line  supplementary materials we present the following data.
In Appendix~A we provide finding charts of all observed galaxies with the
spectrograph long slit position superimposed.
In Appendix~B, plots of 1D spectra for the observed
galaxies are presented. Appendix~C presents the tables with line intensities, derived
physical parameters and the values of 12+log(O/H).

\section[]{SALT observations and data processing}
\label{sec:observing}

For our spectral observations we used the Southern African Large Telescope
\citep[SALT;][]{Buck06,Dono06} in service mode
in the period from May 2019 to April 2023.
Several of the 64 target galaxies were observed from two to three times
in order to increase the S-to-N ratio of the 'strong' lines in the potentially
very low-metallicity objects. The obtained spectra were averaged for the subsequent
analysis.
See Tables~1, 2.   % \ref{tab:journal1, tab:journal2}.
The SALT Robert Stobie Spectrograph  \citep[RSS;][]{Burgh03,Kobul03}
 was used with VPH grating PG0900 with the long slit of 1.5\arcsec\ by 8\arcmin.
This option provides the range from 3600~\AA\ to 6700~\AA\ and the spectral
resolution of FWHM$\sim$6.0~\AA.
For the obtained 2D spectra, we used a binning factor of four for the
spatial scale and factor of two for the spectral coordinate. This gives us a final
spatial sampling of 0\farcs51 pixel$^{-1}$ and spectral sampling of 0.97
\AA\,pixel$^{-1}$ .

Spectral observations with the sufficiently small entrance hole or with the narrow slit
are the subject of the differential atmospheric refraction \citep[e.g.,][]{Filippenko82}.
To minimise its effect on the wavelength-dependent light loss, the long slit
position angle (PA) is recommended to keep close to the direction of the atmospheric
refraction (parallactic angle).
In many cases, this can substantially limit the efficiency of observations.
The RSS is equipped with an Atmospheric Dispersion Compensator (ADC),
what allowed us not to worry about the effect of atmospheric refraction
at arbitrary long-slit PAs. Spectrophotometric standards were observed
during every night as a part of the SALT standard calibrations program.

In order not to repeat literally the description of specifics of SALT observations
and flux calibration, we refer a reader to Section~2 of our recent paper \citep{XMP.SALT}.

The majority of \HII-regions in the observed galaxies are faint and
low-contrast. For the pointing, we used  nearby offset stars.
In most cases, we selected position
angles (PAs, in degrees) of the slit
so that to include a nearby offset star and to cover
a faint \HII-knot in a program galaxy. See the journals of observation in
Tables~1, 2.  % \ref{tab:journal1, tab:journal2}.
They present the main information on each observation:
the dates of observations, exposure times, seeing $\theta$
(in arc seconds) and air mass.

Since many of the observed dwarfs are faint and of low surface brightness,
an attempt  to obtain their independent spectra
can be a problem. To ease independent checks of our data, we present
in Tables~1,~2 the used long slit position angles (PA).
In Figures~A1 -- A4 of the on-line materials, we show
their images with superimposed long slit positions.
All but a few images are taken from the Legacy surveys database \citep{Legacy}
The remaining several images which are unavailable in the Legacy database,
are taken from the ESO Online archive (https://archive.eso.org/dss/dss).

Similar to our previous paper \citep{XMP.SALT}, the SALT science pipelines \citep{Cra2010}
were used for the primary data reduction. For each CCD amplifier, they include
bias and overscan subtraction, gain and cross-talk corrections and finally,
mosaicing. The following long-slit reduction was conducted as described in
the paper by \citet{Kniazev2022}.

\input{tab_jour2.tex}

\section[]{Line measurements and O/H determination}
\label{sec:OH}

The emission line fluxes obtained from 1D spectra were measured as
described in detail in \citet{Kniazev04} and briefly summarised in \citet{XMP.SALT}.
In order not to repeat this description, we refer a reader to these papers.
Very briefly, the procedures include the robust drawing of continuum with
the subsequent use of \mbox{MIDAS}\footnote{MIDAS is an acronym for the
European Southern Observatory package -- Munich Image Data Analysis
System.}-based
programs for determination of parameters of emission lines.

Giving the line fluxes measured,  we utilise
an iterative procedure from \citet{ITL94}, which accounts for the total sight-line
dust extinction and the underlying Balmer line
absorptions originating in the related young stellar clusters.
As a result, it provides the simultaneous estimate of the equivalent width of
absorption Balmer lines $EW(abs)$ and the extinction coefficient $C(H\beta)$.
The relevant equation (1) from  \citet{ITL94}  was used:
$$ I(\lambda)/I(H\beta) = [EW_{e}(\lambda)+EW_{a}(\lambda)]/EW_{e}(\lambda) \times  $$
$$ EW_{e}(H\beta)/[EW_{e}(H\beta)+EW_{a}(H\beta)] \times $$
$$ F(\lambda)/F(H\beta)~exp[C(H\beta)f(\lambda)]~~~~~~~~ (1) $$

Here $I(\lambda)$ is the intrinsic line flux corrected for the overall
extinction (both
in the Milky Way and internal to a particular galaxy or \HII\ region) and the underlying
Balmer absorption, while $F(\lambda)$ is the measured line flux.
Here, as in \citet{ITL94}, $EW_{e}(\lambda)$ are equivalent widths of used emission
lines.
$EW_{a}(\lambda)$ is the adopted value of the underlying Balmer absorptions.
The theoretical Balmer line ratios I($\lambda$)/I(H$\beta$) for Case~B
from \citet{Brocklehurst71} were used for iterative procedure with equation (1).

Following to \citet{ITL94}, this term is used to estimate intrinsic fluxes
only for Balmer emission lines. Similar to this paper, we adopt
 the reddening function $f(\lambda)$ from \citet{W1958}. It is
normalised so that $f(H\beta)$ = 0.  For this $f(\lambda)$,
there is a relation between the excess E($B-V$) and $C(H\beta)$:
E($B-V$) = 0.68 $\times$ $C(H\beta)$.

In the classic (T$_{\rm e}$) method of the oxygen abundance determination, they use
the standard two-zone model and method of \citet{Aller1984}. Since the electron
temperature T$_{\rm e}$ is different in high and low-ionisation \HII\ regions
\citep[e.g.][]{Stasinska1990}, to calculate abundances of ions O$^{++}$ and
O$^{+}$, one needs to know temperature T$_{\rm e}$ in both parts of an \HII\ region.
The estimate of T$_{\rm e}$(O$^{++}$) uses the flux ratio of [O{\sc iii}] lines:
I($\lambda$4363)/I([$\lambda$4959 + $\lambda$5007) and the five-level atom model
\citep{Aller1984}, with the electron density N$_{\rm e}$ derived from the line
ratio of the [S{\sc ii}] doublet: I($\lambda$6717)/I($\lambda$6731). Since in our
spectra this doublet was outside the range, we adopted for further calculations
N$_{\rm e} = $ 10~cm$^{-3}$, a typical electron density in \HII\ regions of dIrr galaxies.
To derive T$_{\rm e}$(O$^{+}$), we used the fit for the relation between T$_{\rm e}$(O$^{+}$)
and T$_{\rm e}$(O$^{++}$) from \citet{ITL94}, based on the models of photoionised \HII\
regions by \citet{Stasinska1990}. This fit is read as follows:

 t$_{\rm e}$(O{\sc ii}) = 0.243 + t$_{\rm e}$(O{\sc iii})[1.031 - 0.184 t$_{\rm e}$(O{\sc iii}])] (2)

Here t$_{\rm e}$ = 10$^{-4}$ T$_{\rm e}$. The resulting rms uncertainty of t$_{\rm e}$(O{\sc ii})
is transferred from the corresponding
rms error of t$_{\rm e}$(O{\sc iii}) as derived with use of the above relation.
The ionic abundances of O$^{+}$/H$^{+}$ and O$^{++}$/H$^{+}$ are calculated
with formulae (3) and (5) from \citet{Izotov2006}.

The major part of \HII\ regions in the studied galaxies are rather faint.
Accordingly, the fluxes of their emission lines are low.
The weak auroral line, [O{\sc iii}]$\lambda$4363~\AA,  which is used in
the 'direct' (T$_{\rm e}$) method
for the determination of the electron temperature T$_{\rm e}$,
was detected only in the minority of our targets. Therefore, for the estimate of O/H
in the remaining galaxies, we applied the semi-empirical method
of \citet{IT07}. However, as was shown in \citet{XMP.BTA}, this method provides
the reliable estimates of T$_{\rm e}$ in rather limited range of parameters of \HII\ regions.
We modified their method, as described in \citet{XMP.BTA}, in order
to expand its applicability to the wider range of physical conditions.

The basement of method of \citet{IT07} is the fitted empirical dependence between
the values of the electron temperature T$_{\rm e}$ and parameter $R_{\mathrm 23}$.
It was derived from the analysis of the grid of models of
\HII-regions in \citet{SI2003}. As the authors find, for the large representative sample
of extragalactic \HII-regions, which cover the whole range of observed O/H, the models approximate
well the apparent relations of the strong line intensities versus EW(H$\beta$).
The parameter $R_{\mathrm 23}$ is the ratio of the sum of fluxes
of strong oxygen lines [O{\sc ii}]$\lambda$3727~\AA,
[O{\sc iii}]$\lambda$4959~\AA, [O{\sc iii}]$\lambda$5007~\AA\ to that
of H$\beta$.
When T$_{\rm e}$ is estimated via $R_{\mathrm 23}$, the rest of the
calculations of the ionic abundances of O$^{+}$/H$^{+}$ and O$^{++}$/H$^{+}$
uses the standard equations of the classic T$_{\rm e}$ method, that is
the same formulae (3) and (5) from \citet{Izotov2006}.

As it became clear later, the good approximation of O/H, derived by this
method, to O/H values, obtained by the direct method, has the limited application.
This is because it was fitted from a sample of \HII-regions with a rather limited range of the
excitation parameter $O_{\mathrm 32}$ (the ratio of line fluxes of [O{\sc iii}]$\lambda$5007
and [O{\sc ii}]$\lambda$3727). This parameter is a proxy of the physical parameter
$U$ - so-called the ionisation parameter, characterising the flux of ionising photons per
unit area of the illuminated gas cloud surrounding the central source. While the original
method of \citet{IT07} works well for the intermediate values of $O_{\mathrm 32}$
($\sim$ 2--6), it leads to the significant systematical bias in the T$_{\rm e}$ and
O/H estimates at both, small and large values of $O_{\mathrm 32}$.
The whole range of $O_{\mathrm 32}$ in known \HII\ regions spreads from $\lesssim$0.3
to $\sim$50.

This effect was examined in \citet{XMP.BTA} on the large sample of \HII-regions
with O/H derived via the direct method. The sample covers the broad range of
parameter $O_{\mathrm 32}$ and O/H. The modified formula was suggested to estimate T$_{\rm e}$,
which includes the dependence of T$_{\rm e}$ on the both parameters, $R_{\mathrm 23}$
and $O_{\mathrm 32}$.  The respective method is called the modified
semi-empirical (mse). The mse method has the moderate internal scatter
in log(O/H) of 0.09~dex, while allows one to eliminate the systematic bias
of the original method of \citet{IT07} for the extreme values of parameter $O_{\mathrm 32}$.

This is crucial for our project, since the typical \HII-regions
in our void galaxy sample, in which the auroral line [O{\sc iii}]$\lambda$4363 was
too noisy or undetected,
have rather low excitation parameter, of $O_{\mathrm 32} \lesssim $ 1.0.
The O/H estimates derived by this method are marked as (mse)
in Column~9 of Table~\ref{tab:prop_summary}.

Recently \citet{Izotov19DR14} suggested a new empirical method, the best
suited for \HII-regions with 'metallicity' 12+log(O/H) $\lesssim$ 7.4~dex.
This uses only the relative fluxes of strong Oxygen lines with respect of
H$\beta$. Namely, their Equation (1) reads as:
$$12+\log({\rm O/H}) = 0.950 \times \log(R_{\mathrm 23} - 0.08\times O_{\mathrm 32}) + 6.805 $$
This relation, calibrated on the large number of \HII-regions with O/H derived
via the direct (T$_{\rm e}$) method, empirically accounts for the large
scatter in the ionisation parameter log(U) (from about --4 to  --1) in various
\HII-regions and, thus, reduces the relatively large internal rms scatter of the other
empirical methods based on the strong oxygen lines, down to only $\sim$0.04~dex.
The latter value of
the rms scatter was obtained in \citet{XMP.BTA} on the sample of $\sim$70 \HII-regions
with the range of 7.0 $\lesssim$ 12+$\log$(O/H) $\lesssim$ 7.5~dex.
However, its applicability is limited only to the range of
12+$\log$(O/H) $\lesssim$ 7.5.
The above relation was slightly modified by \citet{Izotov21} to account \HII\ regions
with the highest observed values of O$_{\mathrm 32}$.

Among our galaxies there are 16 objects falling to this
category. For them, we use this method.
For derived this way value of O/H, we attach index (s)
in Column~9 of Table~\ref{tab:prop_summary}, and denote them hereafter
as O/H(s).

As  shown in \citet{XMP.BTA},
there is a tiny offset in the zero-point of 12+log(O/H)(s) relative to
that of 12+log(O/H)(T$_{\rm e}$), of $\sim$0.01 dex.
In Table~\ref{tab:prop_summary}, we present therefore the values of
12+log(O/H)(s)
with the subtraction of 0.01~dex from the values obtained with the formula from
\citet{Izotov19DR14}. This allows us
to compare them directly with the estimates of 12+log(O/H)(T$_{\rm e}$)
for other galaxies.

The contribution of errors in the strong line fluxes to the related uncertainty
 of 12+log(O/H)(s), $\sigma_{\rm log(O/H)}$, is rather modest.
For the high S-to-N spectra with $\sigma_{\rm R23}$/$R_{\mathrm 23} <$0.02, the
total $\sigma_{\rm log(O/H)}$  is close to the internal error of the method,
namely,  $\sigma_{\rm log(O/H)} \sim$0.04~dex.  For the lowest S-to-N our
spectra of
$\sigma_{\rm R23}$/$R_{\mathrm 23} \sim $0.14,
the error of $\log$(O/H)(s) increases to 0.12~dex.
 Intermediate S-to-N of $R_{\mathrm 23}$ results in
the typical $\sigma_{\rm log(O/H)}$ = 0.05 -- 0.06~dex.

\setcounter{qub}{0}
\begin{table*}
\caption{Observed nearby void dwarfs and new O/H data}
\begin{tabular}{r|l|l|r|r|r|r|l|l|p{3.0cm}} \hline\hline
No.&\MC{1}{c}{Name}&\MC{1}{c}{J2000 Coord}& V$_{\rm orig}$ & V$_{\rm SALT}$&D\dg &B$_{\rm t}$ &$M_{\rm B}$ & 12+$\log$(O/H) & Notes \\
   &              &                  &\kms      &  \kms   &Mpc  &mag  &mag     & $\pm$err.          &       \\
 1 &\MC{1}{c}{2}  &\MC{1}{c}{3}&\MC{1}{c}{4}    &   5     & 6   &   7 &     8  &\MC{1}{c}{9}        &\MC{1}{c}{10}   \\
\hline
\qq&J0015+0104    & J001520.7+010437 &2055\p02  &2141\p15 & 28.8&18.31&--14.32 & 7.14$\pm$0.04 d+s  & prototype XMP \\ %
\qq&ESO294-010    & J002633.4--415120& 113\p04  &  84\p20&\dg2.0&15.56&--10.90 & 7.79$\pm$0.12 mse  &   \\ %  changed after included F(Hg) and change of F(3727)
\qq&PGC004055     & J010822.0--381233& 654\p60  &633\p10  & 6.8 &16.09&--13.09 & 7.60$\pm$0.10 mse+d&    \\ %
\qq&UGC01085      & J013118.9+074716 & 652\p02  &680\p07  &10.0 &16.75&--13.44 & 7.87\p0.11 mse     &    \\ % updated PT05 O/H=7.92 +-.10
\qq&AGC124137$^*$ & J023137.0+093144 & 897\p06  &900\p10  &14.0 &17.98&--12.10 & 7.20\p0.12 s+mse   & low S-to-N \\ %
\qq&ESO199-007    & J025804.1--492256& 630\p05  &623\p15&\dg 6.0&16.50&--12.49 & 7.33$\pm$0.06 s    & aver. on 2 knots \\ %
\qq&PGC1166738$^*$& J030646.9+002811 & 710\p04  &728\p20  &11.0 &18.41&--12.27 & 7.19$\pm$0.04 s    & aver. on 2 knots \\ %
\qq&PGC013294$^*$ & J033556.8--451129& 737\p42  &750\p10&\dg 7.3&16.07&--13.29 & 7.90$\pm$0.06 d    & \\ % 2 HII regions from HST image. PA=-24 D=7.3
\qq&PGC712531     & J033903.0--304921& 839\p123 &692\p10  & 9.5 &18.29&--11.96 & ...                & only faint H$\alpha$           \\ %
\qq&PGC681755     & J033955.0--330309& 747\p26  &705\p15  & 9.5 &16.55&--13.39 & 7.71$\pm$0.11: mse  & low O$_{\mathrm 32}$ $\sim$0.25 \\ %
\qq&ESO359-024    & J041057.5--354951& 847\p05  &815\p30  &10.9 &15.43&--14.80 & 7.44$\pm$0.13 d    & \\ %
\qq&PGC016389     & J045658.7--424802& 662\p06  &710\p12&\dg 7.0&14.46&--14.80 & 7.20\p0.05 s       & $\sim$7.0 in O$_{\mathrm 32}$$\sim$0.2  \\ % HIPASSJ0457-42
\qq&HIPASSJ0517-32& J051721.6--324535& 798\p06  &811\p06  &10.7 &15.96&--14.81 & 7.93$\pm$0.05 d    & aver. 2 knots \\ %
\qq&ESO553-046    & J052705.8--204040& 542\p06  &538\p05&\dg 6.7&14.72&--14.71 & 8.03$\pm$0.03 d    &  \\ % SIGRID32
\qq&PGC138836     & J055735.2+072913 & 428\p04  &407\p15&\dg 5.5&18.40&--11.28 & 7.56$\pm$0.10 mse  &  \\ %
\qq&PGC018431     & J060719.7--341216& 774\p13  &816\p07&\dg 9.6&15.66&--14.34 & 7.65$\pm$0.06 d    & aver. 2 knots  \\ % HIPASSJ0607-34 =6dF  3 knots with O/H_d, make aver.!
\qq&ESO308-022    & J063932.9--404317& 822\p04  &888\p11&\dg 9.4&16.22&--14.03 & 7.34$\pm$0.05 s    &  \\ %
\qq&PGC020125     & J070517.7--583108& 564\p04  &489\p05&\dg 5.3&14.95&--14.16 & 7.37$\pm$0.05 s    & aver. 2 knots \\ % Argo DIG
\qq&6dFJ0935-1348 & J093521.6--134852& 800\p44  &877\p22  & 11.7&16.43&--14.10 & 7.28$\pm$0.07 s    & \\ %  Monoceros void program, DNN=1.94+-0.60
\qq&PGC029033     & J100138.2--081455& 446\p03  &258\p28  &  7.0&15.41&--14.07 & 8.01\p0.15    mse  & V(HI)-V(opt)$\sim$190 \\ % OII, Hbeta, Halpha, OIII \\ %
\qq&PGC041667     & J123307.9--003158& 740\p09  &737\p37  & 10.2&16.50&--13.63 & 7.84$\pm$0.14 mse  & \\ %
\qq&PGC091215     & J123655.0+013654 & 590\p01  &623\p12  &  6.1&16.35&--12.65 & 7.37$\pm$0.05 s    & aver. on 3 knots \\ % 3 EL regions, w.mean O/H_s,c = 7.37
\qq&AGC227970     & J124601.4+042252 & 643\p04  &625\p10  &  9.2&16.00&--13.93 & 7.71$\pm$0.08 d    & \\ % O/H_mse,c=7.79+-.12
\qq&AGC225197     & J124942.1+052922 & 739\p18  &820\p30  & 10.1&18.38&--11.77 & ...                & abs. spectrum \\ %  J1249+0529 gas-rich only Balm. abs M_B=-11.77 Vh(HI,ALFA)=739+-18, W_50=130+-37! (SDSS: abs.680+-35) Vh(abs.SALT)=866 => Vd=864, D=11.8. MB=~12.1
\qq&UGC07983      & J124947.0+035033 & 694\p02  &743\p20  & 10.1&16.11&--14.08 & 8.03$\pm$0.14 mse  &  \\ %
\qq&PGC1289726    & J124959.2+054916 & 618\p01  &600\p10  &  9.1&16.76&--13.17 & 7.68$\pm$0.11 mse  &         \\ %
\qq&AGC227972     & J125024.0+045422 & 650\p04  &647\p20  &  9.1&19.50&--10.46 & ...                & only H$\alpha$ in 2 knots \\ %   V1=430 (W=4.1, S/N~6) V2=619 (W=7.5, S/N~10) Del~V=190 km/s dY12~6" => 0.27 kpc NEED to discuss! +V(Hg-abs)=647
\qq&AGC227973     & J125039.9+052052 & 675\p04  &695\p15  &  9.1&19.50&--10.42 & 7.07$\pm$0.04 s    & aver. on 2 measur. \\ % O/H_s varies from 7.05 to  7.09  and to 7.15
\qq&AGC226122     & J125215.4+042727 & 700\p04  &756\p15  & 10.0&18.33&--11.86 & ...                & only H$\alpha$     \\ %
\qq&PGC1264260    & J125343.3+040914 & 761\p01  &810\p12  & 10.0&17.45&--12.69 & 7.58$\pm$0.08 d    &                    \\ %
\qq&UGC08055      & J125604.4+034846 & 616\p02  &622\p18  &  6.7&15.94&--13.34 & 7.79$\pm$0.11 mse  &                    \\ %
\qq&PGC044681$^*$ & J125956.6--192441& 827\p02  &796\p15&\dg 7.3&17.00&--12.73 & 7.34$\pm$0.08 s    & check earlier data  \\ %
\qq&KKH86         & J135433.4+041440 & 286\p02  &298\p30&\dg 2.6&17.08&--10.14 & ...                & only Bal. absorb.  \\ % PGC2807150
\qq&AGC716018$^*$ & J143048.7+070926 &1365\p04  &1415\p12 & 18.1&18.18&--13.22 & 7.93$\pm$0.14 mse  &                    \\ %
\qq&AGC249197$^*$ & J144950.7+095630 &1809\p05  &1822\p10 & 24.2&18.69&--13.35 & 7.39$\pm$0.06 s    &                    \\ %  Preliminary, took F(Hg) from plot
\qq&HIPASSJ1738-57$^*$&J173842.9--571525&858\p08& 815\p13 &  7.3&16.62&--13.12 & 7.72$\pm$0.11 mse  & 7.77$\pm$0.18 (d)  \\ % O/H_d = 7.77 +- 0.18
\qq&PGC408791     & J202608.8--552950& 893\p67  & 812\p10 &  6.8&16.88&--12.71 & 7.39$\pm$0.05 s    &                    \\ %
\qq&PGC129680     & J210804.8--471942& 860\p33  & 858\p12 &  7.7&15.69&--13.90 & 8.00$\pm$0.10 mse  & aver. 2 knots      \\ %
\qq&PGC1016598$^*$& J213902.9--073443&1283\p02  &1305\p12 & 15.4&18.63&--12.43 & 7.23$\pm$0.04 s    & aver. 2 knots      \\ %  2 close knots at 3.5": give comment
\qq&ESO289-020    & J222111.7--454035& 912\p67  & 892\p06 &  8.8&15.62&--14.17 & 7.59$\pm$0.05 mse  & aver. 3 knots      \\ %  220720 895+-5 dV=+7 220428 876+-6 dV=?
\qq&ESO238-005$^*$& J222230.5--482414& 706\p03  &714\p04&\dg 8.0&15.30&--14.28 & 7.35$\pm$0.04 s    &                    \\ %
\qq&PGC1028063    & J224223.4--065010& 899\p06  & 861\p06 & 10.7&16.02&--14.28 & 7.79$\pm$0.09 d    &                    \\ %
\qq&ESO347-017    & J232656.2--372049& 691\p03  & 698\p06 &  6.5&14.93&--14.20 & 7.82$\pm$0.03 d    & aver. 3 knots      \\ %
\qq&PGC3192333    & J234133.8--353023& 630\p89  & 524\p18 &  6.6&18.86&--10.31 & 7.26$\pm$0.08 d    & 7.18$\pm$0.05 (s)  \\ %  Need to change D according. to new Vh!
\qq&PGC680341     & J234147.5--330841& 510\p89  & 446\p10 &  4.4&16.59&--11.69 & 7.34$\pm$0.04 d    & 7.29$\pm$0.04 (s)  \\ %  Need to change D according. to new Vh!
\qq&ESO348-009    & J234923.4--374622& 647\p02  & 587\p15 &  6.1&14.81&--14.18 & 7.42$\pm$0.05 s    &                    \\ %
\qq&ESO149-003    & J235202.8--523438& 574\p04  &526\p09&\dg 7.0&15.09&--14.20 & 7.66$\pm$0.03 d    &                    \\ %
\hline
\multicolumn{10}{p{17.5cm}}{Table~3 content is described in detail in Sect.~\ref{ssec:results_void}. Here
we give the brief information. Col.~2: target name from NVG.} \\
\multicolumn{10}{p{17.0cm}}{Col.~3: galaxy coordinates adopted from NVG. Col.~4: original radial velocity with its error,
in \kms. Col.~5: SALT radial velocity } \\
\multicolumn{10}{p{17.0cm}}{with its error; Columns~6, 7 and 8: the adopted distance, total blue magnitude and absolute blue
magnitude. Col.~9: derived O/H} \\
\multicolumn{10}{p{17.0cm}}{as 12+$\log$(O/H) and its error, in dex, with indication of the method used: (d) -- the direct
(T$_{\rm e}$) method; (mse) -- the modified} \\
\multicolumn{10}{p{17.0cm}}{semi-empirical method of \citet{IT07}, accounting for the large range of excitation
\citep{XMP.BTA};} \\
\multicolumn{10}{p{17.0cm}}{(s) -- the new empirical strong line O/H estimator of \citet{Izotov19DR14}, with subtracted
0.01~dex to account for a small offset} \\
\multicolumn{10}{p{17.0cm}}{relative to O/H(T$_{\rm e}$). In Col.~10 we show brief notes with more detailed information,
when necessary, presented in Sec.~\ref{ssec:comments_XMP}, \ref{ssec:comments_void}.} \\
\multicolumn{10}{p{17.0cm}}{$^*$ - from list of 60 XMP candidate in PEPK20;  \dg\ marks velocity-independent distances
obtained with the TRGB method at HST. } \\
\end{tabular}
\label{tab:prop_summary}
\end{table*}

\input{tab_mistake.tex}

\section[]{Results of spectral observations and O/H estimates}
\label{sec:results}

\subsection{Void galaxies}
\label{ssec:results_void}

The 1D SALT spectra of the observed void dwarfs are presented in the on-line supplementary materials
in Appendix~B, Figs.~B1--B4.    % \ref{fig:SALT_1Dp1} -- \ref{fig:SALT_1Dp4}.
The measured fluxes of emission lines and their derivatives: EW(abs) --
the adopted equivalent width of Balmer absorption in the underlying stellar continuum,
the extinction  coefficient C(H$\beta$), and the equivalent width of
the H$\beta$ emission line EW(H$\beta$) are presented in Tables~C1-C14 of the on-line
supplementary materials in Appendix~C. Some of the obtained spectra, show rather
strong Balmer absorptions in the UV. In these cases the underlying continuum and
Balmer absorptions were fitted by a model with
the ULySS package \citep[http://ulyss.univ-lyon1.fr,][]{Koleva2009}.
This model model spectrum corrected the flux of H$\beta$ emission
to a first approximation.
In these cases, the EW(abs) derived with the procedure from \citet{ITL94}
at the next step, appears  the residual EW(abs),  since this is mainly
accounted for by the ULySS fitting.
As mentioned in Section~\ref{sec:observing}, the absolute flux calibration
at SALT is rather uncertain. Therefore, we do not provide
 the absolute flux in the emission H$\beta$.

The derived physical parameters --- T$_{\rm e}$ in the
zones of emission of [O{\sc iii}] and [O{\sc ii}], the relative numbers of
ions O$^{+}$, O$^{++}$ and the total abundance of Oxygen
relative to Hydrogen, O/H --- are shown in the bottom of these tables.
As described in Section~\ref{sec:OH},
T$_{\rm e}$(O$^{++}$) is calculated with the direct method when the faint auroral
line [O{\sc iii}]$\lambda$4363 is detected. In the remaining cases it is estimated
with the modified semi-empirical method from \citet{XMP.BTA}, based on the semi-empirical
method of \citet{IT07}. As explained in that section, T$_{\rm e}$(O$^{+}$)
was calculated via the relation (2), adopted from \citet{ITL94}.

Since the [S{\sc ii}]$\lambda\lambda$6717,6731 was outside the range of SALT spectra,
the estimate of electron density N$_{\rm e}$ in the observed \HII\ regions was
impossible. We adopted for further calculations of O/H in the direct and mse
methods N$_{\rm e}$ = 10~cm$^{-3}$ as a typical value for \HII\ regions in
the low-mass late-type galaxies.

For each galaxy, we present the derived parameter 12+$\log$(O/H).
For majority of the obtained spectra, this is derived either
with the direct method (for the twelve objects, where it is applicable),  or
with the empirical strong-line method of \citet{Izotov19DR14} (for 15
the lowest O/H objects). For the remaining 12 objects, with the higher values of
O/H, we apply a modified semi-empirical (mse) method, in which we account
for the dependence of the empirically derived T$_{\rm e}$ on the
excitation parameter $O_{\mathrm 32}$  \citep[see details in][]{XMP.BTA}.

We note that 12+$\log$(O/H), derived with the mse
method, needs a small correction (up to 0.03~dex, depending on the value of O/H)
in order to make its zero-points consistent with that for the
direct T$_{\rm e}$ method. See Appendix in \citet{XMP.BTA} for details.
This correction is already applied in Tables~C1--C15 and the respective value of O/H
 is indicated as (mse,c).
It is reasonable to add a caution on the applicability of mse method for the lowest
excitation \HII\ regions. As discussed in the Appendix of \citet{XMP.BTA},
the fitting formulae for deriving T$_{\rm e}$ via this method, are limited
by the lower values of parameter $O_{\mathrm 32}$. For the range of
12+$\log$(O/H) $>$ 7.5~dex, this lower limit of $O_{\mathrm 32}$ is $\sim$0.4,
while for the range of 12+$\log$(O/H) $<$ 7.5~dex, this limit is $\sim$0.5.
For several our galaxies with the lowest excitation, the used formulae are slightly
extrapolated outside these limits. For these objects, the uncertainty of the derived
value of 12+log(O/H) can be somewhat underestimated.

Finally, where it is suitable, just to demonstrate the consistency of our estimates
with other popular methods, we provide the value of O/H, derived with the empirical
estimator of \citet{PT05} (lower branch), which is based only on the flux ratios of the
strong Oxygen lines.
Their formulae also account for parameter of excitation $O_{\mathrm 32}$, which
enters to their parameter 'P'. Their formulae are obtained via the fitting
of the observational data for \HII\ regions with the direct value of 12+$\log$(O/H).
However, the applicability of their formulae is limited by the range of parameter 'P' of
0.55 and 0.97. This translates to the range of $O_{\mathrm 32}$ of 0.92 to 24.2.
Therefore, for our galaxies with the lowest values of $O_{\mathrm 32}$, we do not show
values of O/H (PT05) due to their systematic shifts.

In Table~\ref{tab:prop_summary} we summarise the adopted parameter 12+log(O/H) and provide
some other parameters of the studied galaxies.
The content of the column is as  follows: Col.~1 -- the number of galaxy,
as it appears in Table~1; % \ref{tab:journal1};
Col.~2 -- the galaxy name as adopted in the NVG catalog, which, in turn,
is mainly from the HyperLEDA database\footnote{http://leda.univ-lyon1.fr};
Col.~3  -- J2000 epoch coordinates; Col.~4 -- the original heliocentric velocity
with the cited error, in \kms; Col.~5 -- the heliocentric velocity with its
error obtained from our SALT spectra;  Col.~6 -- Distance in Mpc. For 13 galaxies
this is measured with the Tip of RGB (TRGB) method. For the remaining objects,
the peculiar velocity correction is used according to the velocity
field from \citet{Tully08} as also adopted in the Nearby Void Galaxies catalog
(PTM19); Col.~7 -- an estimate of the total $B$-band magnitude;
Col.~8 -- the absolute magnitude $M_{\rm B}$, with the adopted
MW extinction correction from \citet{Schlafly11};
Col.~9 -- the value of 12+log(O/H), its 1-$\sigma$ uncertainty and the
method used; Col.~10 -- notes for some of the program objects.

In total, we observed on both programs 47 objects belonging to the NVG (Nearby Void Galaxy)
sample. Also, we observed 15 more galaxies, adopted in the NVG catalog to reside in
the nearby voids. This was done based on their HyperLEDA radial velocities.  On results of
our spectroscopy, they appeared to be not the NVG objects. Two more galaxies were observed in order
to check their radial velocities. One was a potential companion of the void galaxy KKH86, while another
galaxy was in a visual contact with the void galaxy PGC16389.
We devote to these 17 objects  Sect.~\ref{ssec:results_mistaken}.

Nine galaxies (of them, eight -- new) were observed as a continuation of the two
previous papers to search for XMP dwarfs \citep{XMP.SALT,XMP.BTA} among the 60 preselected
void XMP candidates \citep{PEPK19}. Accounting for results on 46 already published
objects from this program and several remaining galaxies in the Northern
hemisphere, observed at BTA, the program is very close to its completion.

\subsubsection{Comments on the most metal-poor galaxies}
\label{ssec:comments_XMP}

{\bf SDSS J0015+0104 = AGC103435.}
This galaxy, residing in the Eridanus void \citep{Pustilnik13,Eridanus}, appeared as an XMP
object in two papers, with the values of 12+log(O/H) from 7.07\p0.06 to 7.03~dex
\citep{Guseva09, Izotov19DR14}. % the SDSS and APO spectra.
This galaxy is one of the prototype objects for the search for void XMP dwarfs. Therefore, it was
important to improve its O/H accuracy.
The line [O{\sc ii}]$\lambda$3727 was either outside the available spectral range \citep{Izotov19DR14}, or
the original  semi-empirical method was used \citep{IT07} for too low excitation parameter O$_{\mathrm 32}$ of
$\sim$0.5 \citep{Guseva09}. Our spectrum is quite similar to that of \citet{Guseva09}, with a
higher S-to-N ratio, since we have even marginally detected a faint [O{\sc iii}]$\lambda$4363.
The main difference with their spectrum is our factor $\sim$1.2 larger relative fluxes of
[O{\sc ii}]$\lambda$3727 and [O{\sc iii}]$\lambda$5007.

{\bf AGC124137 = J023137.0+093144.}  This new XMP dwarf was originally selected as a candidate XMP object
in \citet{PEPK19}. The emission lines of the only compact knot are faint and overlay on the underlying
blue continuum with the Balmer-line absorptions. In Figure~B1  % \ref{fig:SALT_1Dp1}
we show both, the original
spectrum (black) and the result of the subtraction of the SSP model spectrum (blue) after the ULySS
package application. While the S-to-N in the used lines for the strong-line and mse methods is quite low,
both methods give the consistent O/H estimates. Since the uncertainty of the measured flux of H$\gamma$ is
large, we varied it in order to check its effect on the derived C(H$\beta$) and EW(abs) and on the value of
O/H. While the C(H$\beta$) varied from 0.32 up to 0.66--0.69, the respective estimates of 12+log(O/H) (s) varied
from 7.08 to 7.30--7.32~dex. The value of C(H$\beta$) related to the Milky Way extinction is 0.13. Therefore,
we believe that C(H$\beta$) = 0.32, adopted in Table~C2,     %\ref{t:Intens2},
is a more reliable since is a more typical for this type of dwarfs.
However, taking into account the low S-to-N of the data, we adopt for this object in
Table~\ref{tab:prop_summary}, the value
of O/H, which is an average between the two current extreme estimates.
The better quality data will be necessary to qualify
confidently this dwarf as an XMP void galaxy.

{\bf PGC1166738 = J030646.9+002811}. This new XMP dwarf was originally selected as a candidate XMP object
in \citet{PEPK19}. We show in Table~\ref{tab:prop_summary} the value of O/H as an average for two knots.
For the knot with the extremely low value of 12+log(O/H)(s) = 7.16$\pm$0.05~dex, we provide all data in
Table~C3.  % \ref{t:Intens2}.
For the second knot, we obtain the value of 12+log(O/H)(s) = 7.22$\pm$0.05~dex.

In the spectrum  of this XMP knot, obtained on 2019.09.06, we detected unusual transient emission lines
(H$\beta$, H$\alpha$, and [O{\sc iii}$\lambda$5007]), which differed in velocity and strength from those observed
in the two later dates, in December 2019. Since the available data are too limited, the nature of this
transient remains puzzling. Probably, new observations of this knot will uncover a puzzle of this phenomenon.

{\bf PGC016389 = J045658.7--424802 = HIPASSJ0457--42}.
In this blue patchy elongated oval we
have on the slit three different \HII\ regions with varying S-to-N emission lines  and with
substantial contribution of the underlying Balmer absorptions. To correctly subtract this continuum, we
modelled it with the ULySS package as explained in Section~\ref{ssec:results_void}. In Figure B1 (Appendix B,
supplementary material)  we present the 1d
spectrum of the most metal-poor region (a). The respective line fluxes and derived physical parameters for
this region are presented in Table C4 (Appendix C, supplementary material). The very low value of
12+log(O/H) (s) = 6.98$\pm$0.05~dex
derived with the method of \citet{Izotov19DR14} is practically insensitive to the adopted value T$_{\rm e}$ (with
the range of 18.2 KK to 26.2 kK, estimated either via semi-empirical method of \citet{IT07}, or with the modified
method (mse) from \citet{XMP.BTA}.

The main problem with this (as well as with 12+log(O/H) (mse)) estimate, is
the very low excitation of this region. The parameter O$_{\mathrm 32} \sim$0.2 is 2.5 times smaller than
its lower boundary for the sample of \HII\ regions with the direct O/H, which is used to derive the empirical relation
in \citet{Izotov19DR14}. Therefore, we should treat the derived very low value of O/H with a caution and rather as an
indicative one. In two other regions, 'b'  and 'c', we derived 12+log(O/H) (s) of 7.20$\pm$0.05~dex and 7.31$\pm$0.09~dex,
respectively. While for region 'c', O$_{\mathrm 32}$ is very small, of $\sim$0.1, for region 'b', O$_{\mathrm 32} \sim$0.46.
Having all this information in hands, we currently adopt its 12+log(O/H) (s) = 7.20$\pm$0.05~dex. Since the galaxy
shows multiple SF regions, we hope that the follow-up high S-to-N spectroscopy will allow one to better determine
the range of metallicities in this dwarf.

{\bf AGC227973 = J125039.9+052052}.  This new faint XMP dwarf, with the value of
12+log(O/H) = 7.07~dex, appears to be one of the lowest metallicity LV dwarfs. It is similar on many
parameters to the other void XMP galaxy, the Leoncino dwarf (AGC198691 = J0943+3326)
\citep{Hirschauer16}. They have similar distances and atomic gas masses, very close
absolute magnitudes and metallicities and amplitude of \HI\ gas motions. The main difference
is the a much higher excitation of \HII\ region in AGC198691 that allows to determine its
metallicity via the direct method and to address the issue of the primordial
 Helium \citep{Aver2022}.

{\bf PGC044681 = J125956.6--192441}.  For this dwarf, we already obtained the SALT
spectrum presented in \citet{XMP.SALT}. For those rather noisy data for the used emission lines,
we found its 12+log(O/H)(s) = 7.22$\pm$0.09~dex, with the upward correction
by 0.02~dex relative to that paper, due to the updated zero-point correction for \citet{Izotov19DR14}
method, as described in Sect.~\ref{sec:OH}. Since that time, a much better quality images
of this dwarf appeared in the Legacy surveys and the Hubble Legacy Archive. We use them to select
the different slit position for the repeat observation. The new value of 12+log(O/H)(s) =
7.34$\pm$0.08~dex, is, from the one hand, marginally consistent with the first measurement.
On the other hand, since it probes the other \HII\ region, it does not exclude the lower
value for that region.

\subsubsection{Comments on the other individual void galaxies}
\label{ssec:comments_void}

{\bf ESO308-022 = J063932.9--404317}. The slit was positioned far from the main
dwarf galaxy body, on the outer SF region well seen at the HST image. Its radial
velocity differs by $\sim$60~\kms\ from that of the main galaxy derived from its \HI\
emission  \citep{HKKE2000}. Since the width of \HI\ profile,
W$_{\mathrm 20} \sim$ 70~\kms, one can think that the studied here SF region can belong
to ESO308-022, or alternatively, is a smaller gas-rich companion in which the
current SF episode was triggered by the tidal interaction with a more massive neighbour.
Probably \HI\ mapping of this galaxy will clear up the nature of this star-forming gas blob.

{\bf PGC029033 = J100138.2--081455}.
For this galaxy, we detected unusually large difference between the \HI-line velocity
of V(\HI) = 446$\pm$1.5~\kms\ \citep{HKK2003} and our emission line velocity V(opt) = 258$\pm$28~\kms\
for the slit position crossing the W edge of galaxy body. The rotation velocity of the galaxy is rather
small as follows from the width of \HI\ profile W$_{\mathrm 0.5}$ = 49~\kms\ \citep{HKK2003}.
While the optical morphology of this dwarf does not look disturbed, this very large difference
in radial velocities of the atomic and ionised gas hints on its non-equilibrium state, probably
caused by a recent interaction or a minor merger. The gas metallicity of this dwarf, despite bearing
rather large uncertainties, also looks rather enhanced in comparison to the similar luminosity  void dwarfs.
Also, it is not clear which of the two velocities reflect the systemic velocity of the galaxy,
which is used for the
distance determination. For the moment, we use the \HI-based velocity.
So, the more detailed study of \HI\ gas kinematics can shed light on the recent processes in this object.

{\bf PGC1016598 = J213902.9--073443.}  In this galaxy the emission on the slit is splitted onto two close
($\sim$7~pixels, or $\sim$3.5~arcsec, in between)
knots (W and E) with quite different flux ratios of [O{\sc iii}]$\lambda\lambda$4959,5007 lines to H$\beta$.
The half-widths of their extent along the slit are close to the distance between the knots,
so that there is a substantial overlapping of their light if we simply extract spectrum of each knot.
To obtain the more reliable line strengths in the E and W knots, we undertook a two-gaussian
fitting along the slit for all lines of interest.
Despite the knots show quite different excitation parameter O$_{\mathrm 32}$ (1.9 versus 0.44),
their parameter 12+log(O/H) (s) is close within rather small uncertainties (7.26\p0.05~dex and 7.19\p0.06~dex,
respectively). Therefore, we adopt their average value, 7.23\p0.04~dex.

\subsection{Improved velocities and distances for void galaxies}
\label{ssec:improved_velocities}

For several void objects from Table~\ref{tab:prop_summary}, the HyperLEDA
radial velocities have moderate to low accuracies (60 --120~\kms, or for related distances,
$\sim$0.9--1.8~Mpc). For objects in the LV and its environs, this results in the
substantial uncertainties in both the relative positions to the nearby neighbours and
to their distance-dependent parameters. In Table~\ref{tab:prop_summary} we give for all objects
the distances and related M$_{\rm B}$ based on NVG (HyperLEDA) data.
Here we briefly summarise
galaxies, for which we obtained a better accuracy radial velocities.

{\bf PGC712531 = J033903.0--304921.} V$_{\rm hel}$ changed from 839 to 692~\kms\ and the
related D -- from 10.95 to 9.3~Mpc.% Vd from 799 to 652

{\bf 6dFJ0935216--134852.} V$_{\rm hel}$ changed from 800 to 877~\kms\ and the related
D -- from 11.7 to 12.7~Mpc. % Vd from 853 to 930

{\bf PGC408791 = J202608.8--552950.} V$_{\rm hel}$ changed from 893 to 812~\kms\ and
the related D -- from 7.5 to 6.4~Mpc. % Vd from 547 to 466

{\bf PGC3192333 = J234133.8--353023.} V$_{\rm hel}$ changed from 630 to 524~\kms\ and
the related D -- from 6.6 to 5.2~Mpc. % Vd from 485 to 379

{\bf PGC680341 = J234147.5--330841.} V$_{\rm hel}$ changed from 510 to 446~\kms\ and
the related D -- from 4.4 to 3.5~Mpc. % Vd from 322 to 258

\subsection{Mistaken objects}
\label{ssec:results_mistaken}

In Table~\ref{tab:wrong} we summarise the results of observations for
15 objects, which appeared to have wrong radial velocities in HyperLEDA
and/or in the original papers. Due to these errors, they reside far from
the distances adopted in the PTM19, and, hence, should be excluded from the NVG sample.
The two exceptions in this Table are the galaxies [MU2012]J1355+04B and PGC016383.
For the former, the information on its radial velocity was absent.
\citet{MU} suggested that this is a fainter companion of the known LV dwarf KKH86.
For the latter, its known radial velocity
placed the galaxy far outside the distance of 25 Mpc, in which the NVG sample
is picked-up. We obtained its independent value just because positioned the
long slit at the void galaxy PGC016389 so that it also crossed PGC016383.
See Sect.~\ref{ssec:comments_wrong}.

We show 1D spectra of all these objects in the on-line supplementary materials
in Appendix~B, Fig.~B5--B6  %\ref{fig:SALT_1Dp4,fig:SALT_1Dp5},
and present the related data in Table~\ref{tab:wrong}. The Table   % Table~\ref{tab:wrong}
includes the following information. Column 2 -
the name of the object adopted from HyperLEDA. Column 3 - its J2000 coordinates.
Column 4 - the original heliocentric velocity from HyperLEDA used to assign the object
to the NVG sample. Column 5 - the heliocentric velocity on the results of the SALT
spectroscopy. For seven very distant objects we give their redshifts instead of radial velocity.
Column 6 - B-band magnitude adopted from HyperLEDA. Column 7 - estimated absolute B-band magnitude
based on parameters in Columns 5 and 6. Column 8 - brief notes. For several objects,
we present below more detailed comments.

\subsubsection{Comments on individual mistaken objects}
\label{ssec:comments_wrong}

{\bf PGC016383=J045701.2--424803}. This known background galaxy looks to be in contact with
  the unrelated void galaxy PGC016389 from Table~\ref{tab:prop_summary}. The SALT slit was positioned
to cross both galaxies. We did not expect to find something new. However, the E+A  spectrum
of PGC016383 shows the radial velocity of
 V$_{\rm hel}$ = 2953~\kms, about 500~\kms\ smaller than the adopted value in HyperLEDA.

\begin{figure}
\includegraphics[width=6.0cm,angle=-90,clip=]{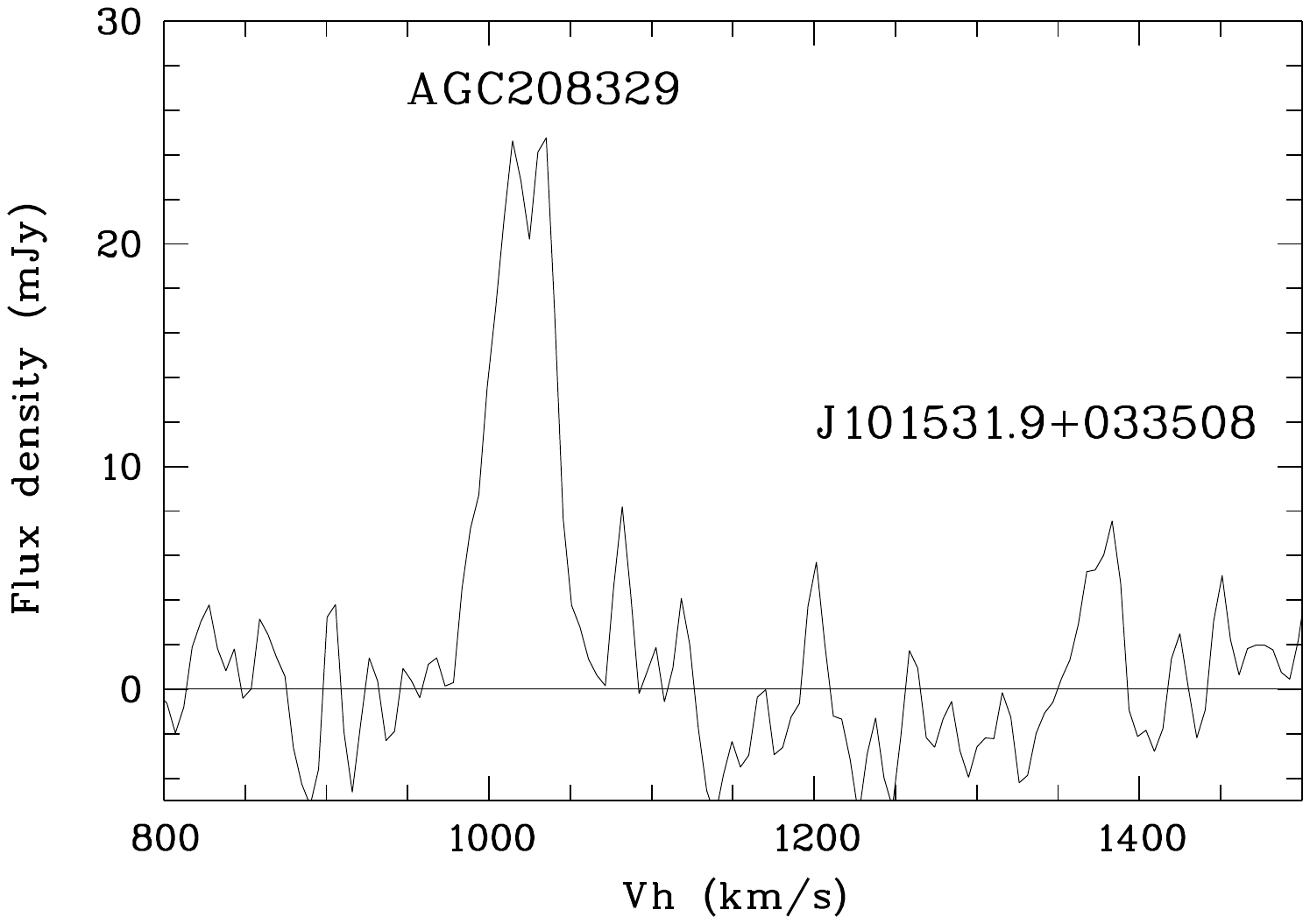}
\caption{ALFALFA \HI\ profile of AGC208329 at V$_{\rm hel}$ = 1018~\kms\ and a
fainter object at V$_{\rm hel} \sim$ 1370~\kms. The latter is close to that of
the optical velocity of SDSSJ101531.88+033508.4 measured in this work.
}
\label{fig:AGC208329_HI}
\end{figure}

{\bf AGC208329 and SDSSJ101531.9+033508}. This blue irregular galaxy is identified in ALFALFA \citep{ALFALFA}
with \HI-source at position J101528.4+033544 with the catalog velocity V$_{\rm hel,HI}$ = 1018$\pm$2~\kms.
The velocity determined on two \HII-regions, V$_{\rm hel,opt} \sim$ 1370~\kms, drastically differs
from that V$_{\rm hel,HI}$. Due to this inconsistency, we have checked the original data of this \HI-source
as presented in the ALFALFA database. Indeed, there is a rather strong source at V$_{\rm hel,HI}$ = 1018~\kms\
(Fig.~\ref{fig:AGC208329_HI}). In addition,
there is a faint, marginally-detected source with F(\HI)$\sim$0.2~Jy~\kms\ at V$_{\rm hel,HI} \sim$ 1370~\kms.
We identify this \HI\ source with the star-forming dwarf, for which we obtained the same radial velocity of
the ionised gas.

The optical galaxy has no alternative names in HyperLEDA.
However, it is identified with two SDSS objects: J101531.88+033508.4 -- for its 'centre', and
J101531.32+033508.9  -- for the blue compact knot at the W edge. Both objects were on the SALT slit.
They have rather similar spectra and the close radial velocities. This SDSS dwarf is seemingly associated
with several dwarf galaxies in the group of the Sa galaxy NGC3169 (J101414.8+032759) with
M$_{\rm B}$ = -20.6~mag, at V$_{\rm hel}$ = 1232~\kms.
Therefore, this optical SDSS object should be excluded from the NVG sample.

As for the nature of AGC208329 itself and an associated with it 'galaxy', it remains
puzzling, since no other potential optical counterparts are visible in the Legacy surveys
colour images within the five-arcmin radius.
One can suggest that the optical counterpart of AGC208329 is a very low SB void dwarf which falls close
to light sight of a much brighter dwarf SDSSJ101531.88+033508.4, for which we obtained optical spectrum.
For the typical ratio of M(\HI)/L$_{\rm B} \sim$ 1, for AGC208329 with F(\HI)= 1.15~Jy~\kms, one
expects for its counterpart a galaxy with B$_{\rm tot}$ = 16.9~mag, 2.5~mag brighter than that
of SDSSJ101531.88+033508.4. To be invisible, this counterpart should be comparable on the total magnitude
with this SDSS dwarf (and hence, to be very gas-rich, with M(\HI)/L$_{\rm B}$ $\sim$10) and
to have a much larger angular extent.
Probably, the future very deep images of this region will detect such an unusual LSB void dwarf.

{\bf [MU2012]J1355+04B = J135429.5+041237.} This blue irregular galaxy was suggested by \citet{MU}
as a counterpart for a nearby void dwarf KKH86 with V$_{\rm hel}$ = 220~\kms\
(see Table~\ref{tab:prop_summary}). Our SALT spectrum shows this is a dwarf
(M$_{\rm B} \sim$ --14~mag) emission-line galaxy at the D $\sim$ 66~Mpc. The strong Oxygen lines
in its \HII\ region allow one to estimate its Oxygen abundance via the 'mse' method described above.
This gives the value of 12+log(O/H) = 7.56$\pm$11~dex.

{\bf PGC264615 = J161354.9--721446, not HIPASS J1614-72.} This blue galaxy was suggested as
an optical counterpart for the faint \HI-source HIPASS J1614-72 by \citet{Kilborn2002}. It is separated from
the Parks radio position (J161422.45-721554.9, with an accuracy of $\sim$1.9~arcmin) by 2.4~arcmin.
For the \HI\ radial velocity of V$_{\rm hel}$(\HI) = 383~\kms, the  distance to this object, derived via
the kinematical model of \citet{Tully08}, is only $\sim$2.5~Mpc. Hence, this \HI-source belongs to a few per cent
of the nearest NVG sample galaxies. Due to the poor coverage of this sky region by the imaging surveys,
its real optical counterpart is still waiting for identification. PGC264615 itself appears a distant SF galaxy
with the redshift of $\sim$0.07. NED (NASA/IPAC Extragalactic Database)
gives a half-dozen galaxies with close redshifts within the radius of 1~degree.

{\bf PGC163318=J212935.2--401653.} The target galaxy looks like as a small disc
with a brighter central core. An off-set red 'star' is situated at $\sim$5~arcsec to NE.
Both objects on the slit display rather similar spectra, with the prominent
absorption features slightly shifted one relative to the other. The non-shifted strong absorption at
the edge of the spectrum is the telluric O$_{\mathrm 2}$ band at $\approx$$\lambda$6870~\AA.
The other prominent absorption doublet at $\sim$$\lambda$6270~\AA\ shows the shift and thus
should be intrinsic for both objects. The most probable identification of this doublet
is  Na{\sc I} D1,D2 (5889.95 and 5895.92~\AA) at the redshift of z = 0.0645. Then 'red' star
appears to be an E-galaxy  with the radial velocity $\sim$450~\kms\ lower than that of the target object.
The fainter absorption features in the spectra of the both galaxies are well consistent on the
wavelengths with the absorption lines visible in spectra of K7-M0 dwarfs from templates in paper
by \citet{Kesseli2017}.
Probably these two galaxies comprise a group with a brighter E galaxy at $\sim$30~arcsec to SW.

\section[]{Discussion}
\label{sec:discussion}

\begin{figure*}
\includegraphics[width=12.0cm,angle=-90,clip=]{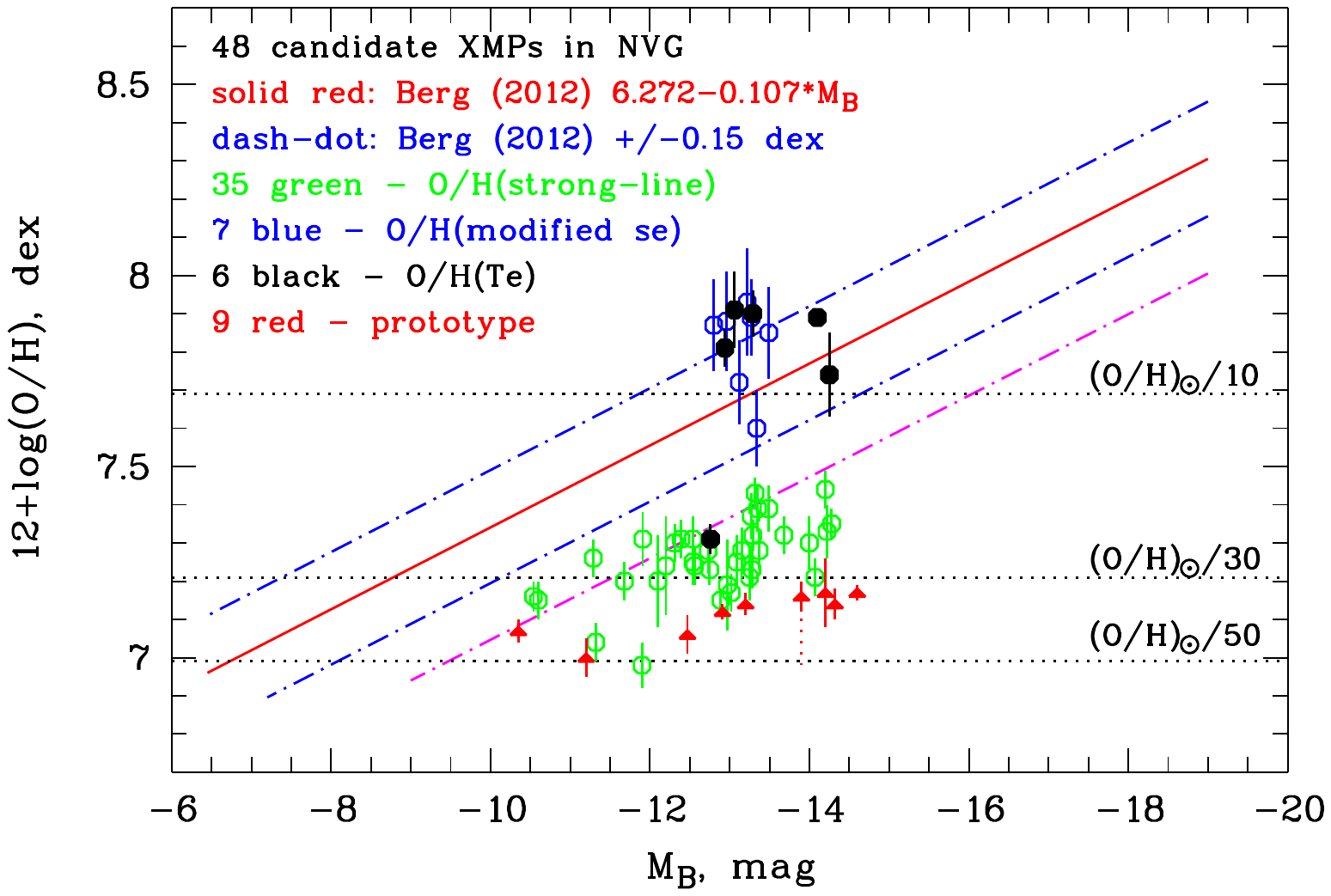}
\caption{Positions of 48 void dwarfs from NVG galaxies selected in PEPK20 as candidate
XMP galaxies. The solid red line is the reference relation derived for a sample of the LV late-type
galaxies by \citet{Berg12}. Two blue dash-dotted lines at $\pm$0.15~dex show the rms scatter of the reference
sample. Magenta dash-dotted line runs at --0.30~dex (--2~rms) from the reference relation.
Nine prototype objects (red triangles) are known from the earlier studies of XMP void dwarfs
summarised in PEPK20. Five of them have O/H determined with the direct method, while the rest four --
with the strong-line method of \citet{Izotov19DR14}. The vertical dashed line at M$_{\rm B} \sim -14.0$
shows the range of O/H for DDO68 from \citet{DDO68_OH}.
}
\label{fig:ZvsMB}
\end{figure*}

As described in the Introduction, this work includes the observational results
on the ionised gas metallicity from two different projects: a) the search for
new XMP dwarfs among galaxies in the NVG sample, and b) the unbiased study
of all void galaxies in the LV, that is about 260 the nearest galaxies of the whole
NVG sample of 1354 objects.  We summarise the respective results below.

\subsection{Search for XMP dwarfs in the NVG sample}
\label{sec:XMP_stat}

Altogether, it was selected 60 NVG sample galaxies with M$_{\rm B} > -14.3$, which on the other properties
were similar to the prototype XMP void dwarfs \citep{PEPK19}. For 46 of them, the spectral  data  were
obtained either with SALT, or with BTA. The related results were published in \citet{XMP.SALT,XMP.BTA}.
In this paper, we present the results of observations for 8 new objects and for one already published
dwarf and briefly summarise the preliminary
statistics of such unusual dwarfs relative to the more common void galaxy population.
% 6 of the above 60 XMP candidates remain in the BTA celestial zone.

Of the preselected 60 NVG XMP candidates, to date, total 54 are observed and presented
%  20 in BTA1-paper (16 w/OH, 1 common with SALT-paper). total 26 - SALT1. 2 -wrong. 2 - only Ha
%  22 w/OH. Total: 37 w/OH. SALT2: 8 new w/OH,   5 unpublished BTA w/OH +1 only Ha
in \citet{XMP.SALT,XMP.BTA} and in this paper. Six more objects remain in the BTA part
(paper in preparation). Two of these 60 candidates appeared galaxies with wrong velocities \citep{XMP.SALT}.
One preselected galaxy appeared too Southern, so that it can not be observed with SALT.
Of the remaining 57 NVG objects, for 50 we were able to derive the parameter 12+log(O/H).
For 11 of these 50, the parameter 12+log(O/H) appears extremely low,
in the range of 6.98 to 7.21~dex. 23 more dwarfs of this {\it candidate} XMP dwarf sample,
have 12+log(O/H) in the range 7.23 -- 7.39~dex,
that corresponds to the range of Z\sunn/30 $\lesssim$ Z(gas) $<$ Z\sunn/20.
That is $\sim$20~percent of that preselected candidates appear XMP galaxies with
Z(gas) $<$ Z\sunn/30. The additional $\sim$40~percent of the preselected objects
appear very-metal poor, with Z\sunn/30 $<$ Z(gas) $<$ Z\sunn/20.

The total number of the {\it nearby} void XMP dwarfs comprises about 20, including the
prototype XMP BCG IZw18 and discovered several decades later, well-known galaxies DDO68,
with 12+log(O/H) in the range 6.98 -- 7.3~dex  \citep{DDO68,IT07,DDO68_OH},
J0926+3343 with 12+log(O/H) = 7.12~dex \citep{J0926},
and Leoncino dwarf (J0943+3326), with 12+log(O/H) = 7.07~dex \citep{Aver2022}.

In Figure~2, %\ref{fig:ZvsMB},
we show how the NVG galaxies, observed as the preselected XMP candidates,
sit on the diagram 12+log(O/H) versus M$_{\rm B}$. The relation between these parameters
was derived by \citet{Berg12} for the reference sample of the LV late-type galaxies
with the well-known O/H and distances. We draw this linear relation and two parallel lines showing
the value of the rms scatter of this sample ($\pm$0.15~dex) around the linear regression.
The pink dot-dashed line is drawn at the 2~rms distance (that is --0.3~dex) below the
reference linear relation. The galaxies with O/H below this line, deviate systematically
downwards from the reference relation.

We use O/H estimates derived with the three different methods, discussed above, depending on
the situation. When the auroral line [O{\sc iii}]$\lambda$4363 is detectable, we use the direct method
(black symbols).
Otherwise, we use the strong-line method of \citet{Izotov19DR14} for the range of
12+log(O/H) $\lesssim$7.5~dex (green symbols) and the mse (modified semi-empirical) method
for the higher O/H (blue symbols). Some  systematics between the data, obtained with the different
methods of the O/H estimate, are basically due to the selection effects.

The appearance of the faint auroral line [O{\sc iii}]$\lambda$4363 correlates with the value
of ionisation parameter log(U) or its proxy, the excitation parameter O$_{\mathrm 32}$, and metallicity.
The former, in turn, are related with the strength of SF 'burst' and its age. Since in the unbiased
sample of void galaxies, the probability to catch a young and strong burst is rather small, this translates
to the low incidence of [O{\sc iii}]$\lambda$4363-line-detected objects.
Similarly, since
we apply either the 'strong-line' or 'mse' methods in the ranges of 12+log(O/H) $<$7.5 or $>$7.5~dex,
the average values of O/H for these two methods show, by definition,  the substantial difference.

\subsection{Intermediate results on metallicities of the LV void galaxies}
\label{sec:LV_stat}

The next steps on the spectroscopy of the NVG dwarfs are conducted in the framework of
the on-going project of the unbiased study of all void galaxies in the LV.
It also finds new very metal-poor objects albeit less frequently.
The smaller incidence of the newly-found XMP dwarfs in this project is a natural
consequence of the unbiased character of the study of the whole NVG sample in the LV. This
is also a reflection of the relative rarity of XMP objects.
In this paper, we add two more LV XMP dwarfs, with 12+log(O/H) = 7.07 and 7.20~dex,
and  7 new dwarfs with 12+log(O/H) $\lesssim$ 7.39~dex.  A couple of new XMP dwarfs within the
LV, which were not preselected as the mentioned above XMP candidates, were also found at the
Northern hemisphere with BTA (paper in preparation).

The unbiased approach in the study of void objects allows one to address their diversity,
including the scatter of gas metallicity for the similar global parameters, such as luminosity or
stellar mass.
To date, we collected data on the gas metallicity in the LV void galaxies
for about a hundred objects. They include both, our own results from this
paper, some of the LV objects appeared in the other our published samples,
as well as about two dozen objects with known O/H, found in the literature.
Despite this number comprises less than a half of the whole LV void sample
(the updated number of 243 objects), it is useful to have a first look on the available data.

For discussion of the intermediate results on the LV sample,
we take at the moment only the data with the direct and strong-line O/H estimates.
The mse estimates of O/H from this and the previous papers will be included in the next
publication, along with the earlier data. The latter need a more careful reanalysis,
including those where we used the semi-empirical method of \citet{IT07}.

\begin{figure*}
\includegraphics[width=12.0cm,angle=-90,clip=]{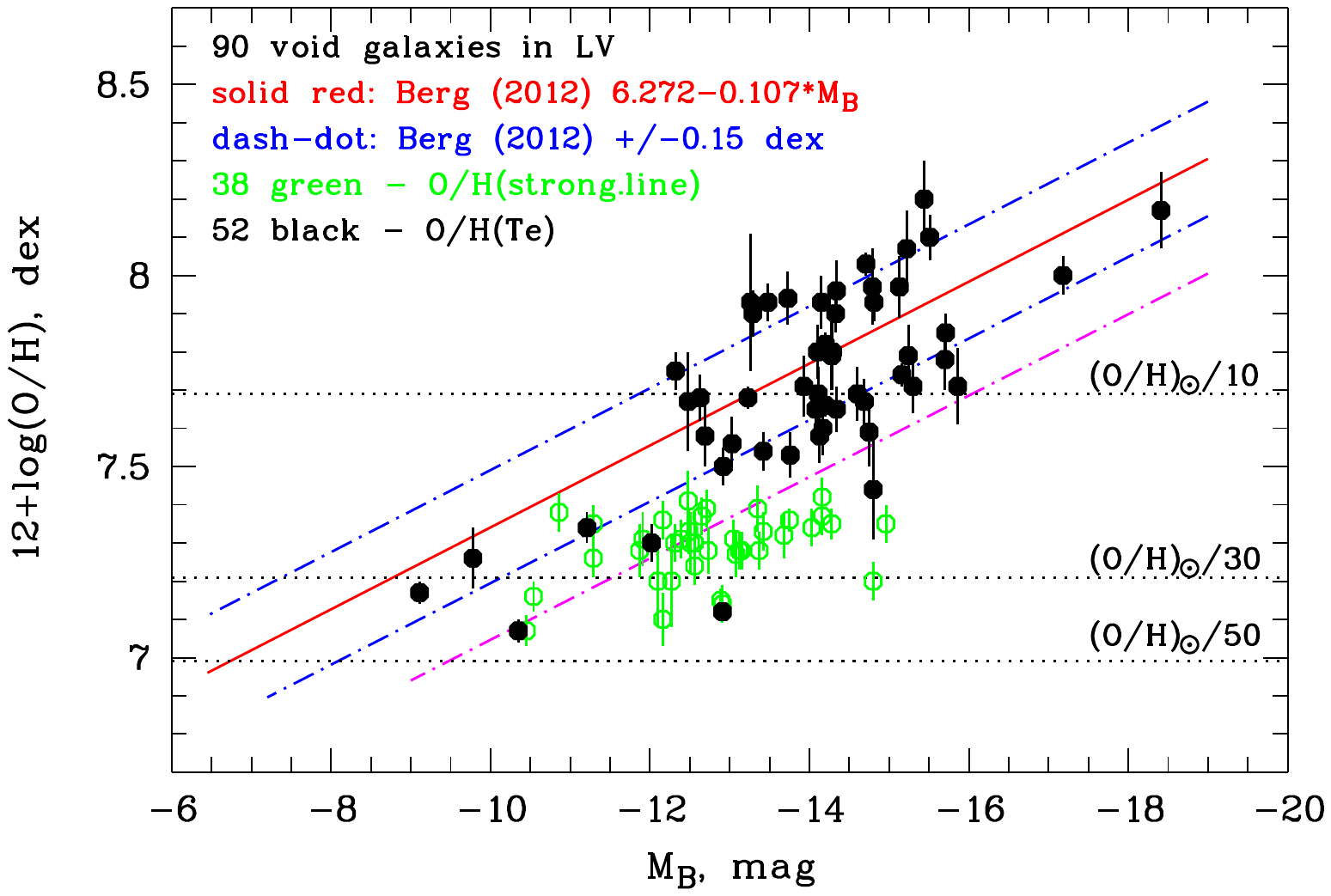}
\caption{Distribution in the diagram '12+log(O/H) vs M$_{\rm B}$' of the 90 LV void
galaxies with known O/H, derived either by the direct (T$_{\rm e}$), or by the
 'strong-line' \citep{Izotov19DR14} methods. The lines on the plot are the same as in Figure~2.
The data  include results
from this paper (25 objects) and from our published papers \citet{PaperVII, XMP.SALT, XMP.BTA}
(31 objects), as well as 12 unpublished BTA objects and 22 objects from the literature
\citep{ITL94,ITL97,IT07,ITG12,vZee97,vZee06a,vZee06b,VV2,Kniazev05,Berg12,Skillman13}.
}
\label{fig:ZvsMB_LV}
\end{figure*}

The plot with the intermediate results for the LV void sample is presented
in Figure~\ref{fig:ZvsMB_LV}. The lines for the reference sample are
the same as in Figure~\ref{fig:ZvsMB}.

One of the features, which distinguishes this void galaxies sample from, e.g.,
the sample of Lynx-Cancer void, is its heterogeneity in the context of their
distances to the nearest massive galaxies. As described in PTM19, the NVG
galaxy sample was selected of objects falling within the empty spheres comprising
the voids themselves. About 20 per cent of this way selected galaxies appeared to
reside closer than two~Mpc to the bordering massive galaxies. They were conditionally
named outer void galaxies,
while the great majority of galaxies ($\sim$80 per cent) with the D$_{\rm NN}$ (nearby
luminous neighbour) distances larger than 2 Mpc were assigned to the inner
void galaxies.

\begin{figure}
\includegraphics[width=6.5cm,angle=-90,clip=]{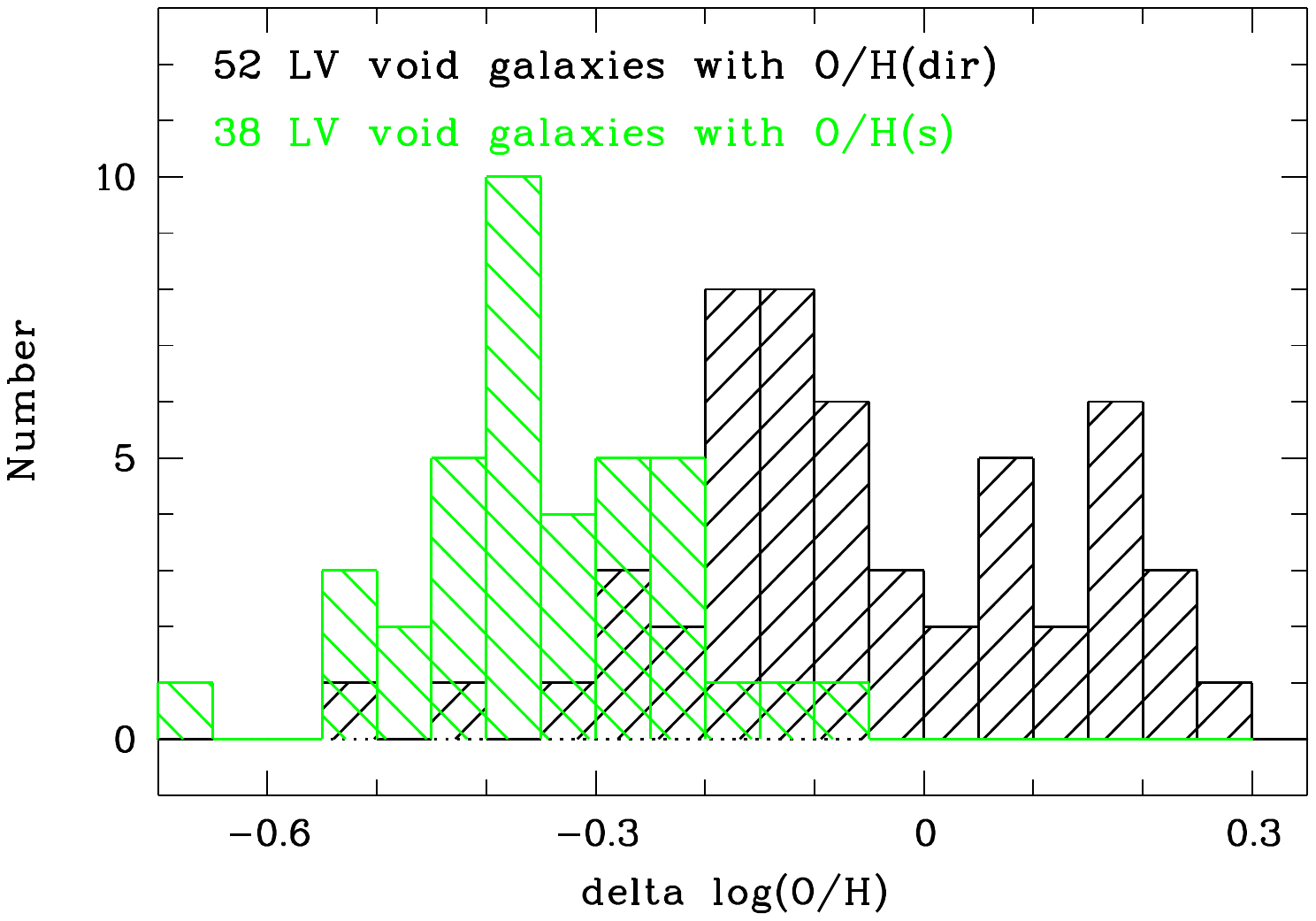}
\caption{The same data as in Figure~\ref{fig:ZvsMB_LV}. The histograms show the  differences of measured 12+log(O/H)
for the LV void galaxies and those expected from the reference relation from \citet{Berg12}. Black hatched
histogram is for O/H(T$_{\rm e}$), while the green histogram is for O/H derived with the strong-line
method of \citet{Izotov19DR14}.
%28 of 90 void galaxies have the parameter 12+log(O/H) reduced more than by 0.30~dex
%relative to the reference relation.
}
\label{fig:hist1}
\end{figure}

As one can see in Figure~\ref{fig:ZvsMB_LV}, the scatter of the LV void galaxies (filled and
empty octagons with error bars) on
log(O/H) at a fixed M$_{\rm B}$, appears significantly larger than for the reference sample of
\citet{Berg12}. For the latter sample, the scatter is mostly limited by the blue dot-dashed
parallel lines ($\pm$1~rms scatter).
While the sizeable part of void galaxies falls within the log(O/H) range of the reference sample,
about one third of the void objects have a substantially reduced metallicity. In particular,
in Figure~\ref{fig:hist1} one can see that 28 of 90 LV void galaxies have the log(O/H), reduced by
more than 0.30~dex (two rms) relative to their values expected from the reference relation of \citet{Berg12}.

The majority of void objects with the substantially reduced metallicity in Figure~\ref{fig:ZvsMB_LV}
belong to galaxies with very low O/H, namely with 12+log(O/H) $\lesssim$7.4. Of them, the great majority
have values of O/H derived with the strong-line method of \citet{Izotov19DR14}. In Figure~\ref{fig:hist1} we show
histograms of the difference 12+log(O/H)(observed) -- 12+log(O/H)(reference, M$_{\rm B}$), separately for
O/H obtained by the direct method (black) and O/H, derived with the strong-line method of \citet{Izotov19DR14} (green).
The mean difference for O/H(T$_{\rm e}$) is \mbox{--0.049}~dex, with rms=0.175~dex. For O/H(s), the mean is --0.346~dex,
with rms=0.116~dex.

\begin{figure*}
\includegraphics[width=12.0cm,angle=-90,clip=]{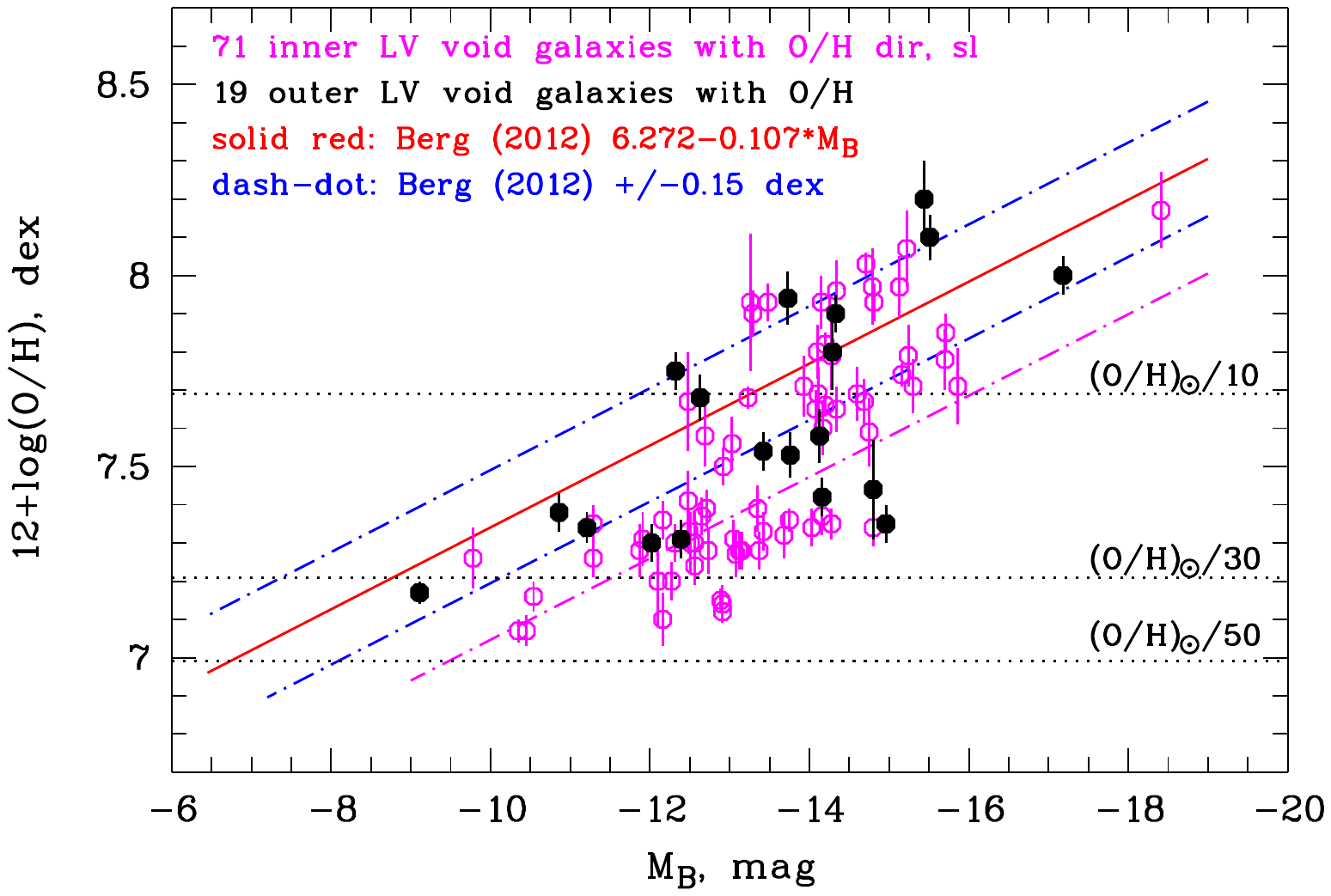}
\caption{The same data as in the previous figure, but divided according to their
distribution within voids.
19 'outer' void galaxies are shown in black. 71 'inner' void galaxies are coloured
with magenta.
See a more detailed discussion in the text.
}
\label{fig:ZvsMB_in+out}
\end{figure*}

\begin{figure*}
\includegraphics[width=6.0cm,angle=-90,clip=]{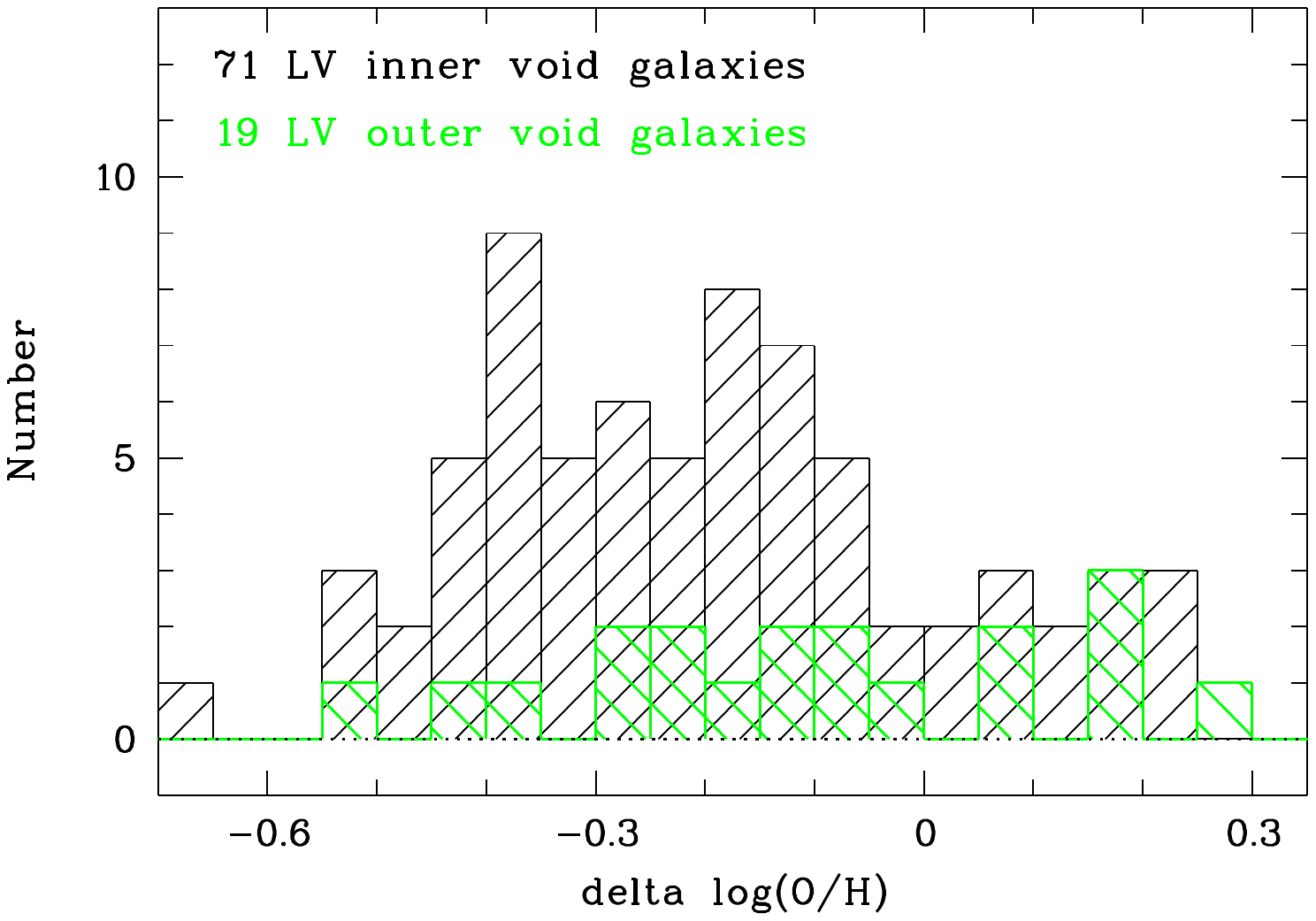}
\includegraphics[width=6.0cm,angle=-90,clip=]{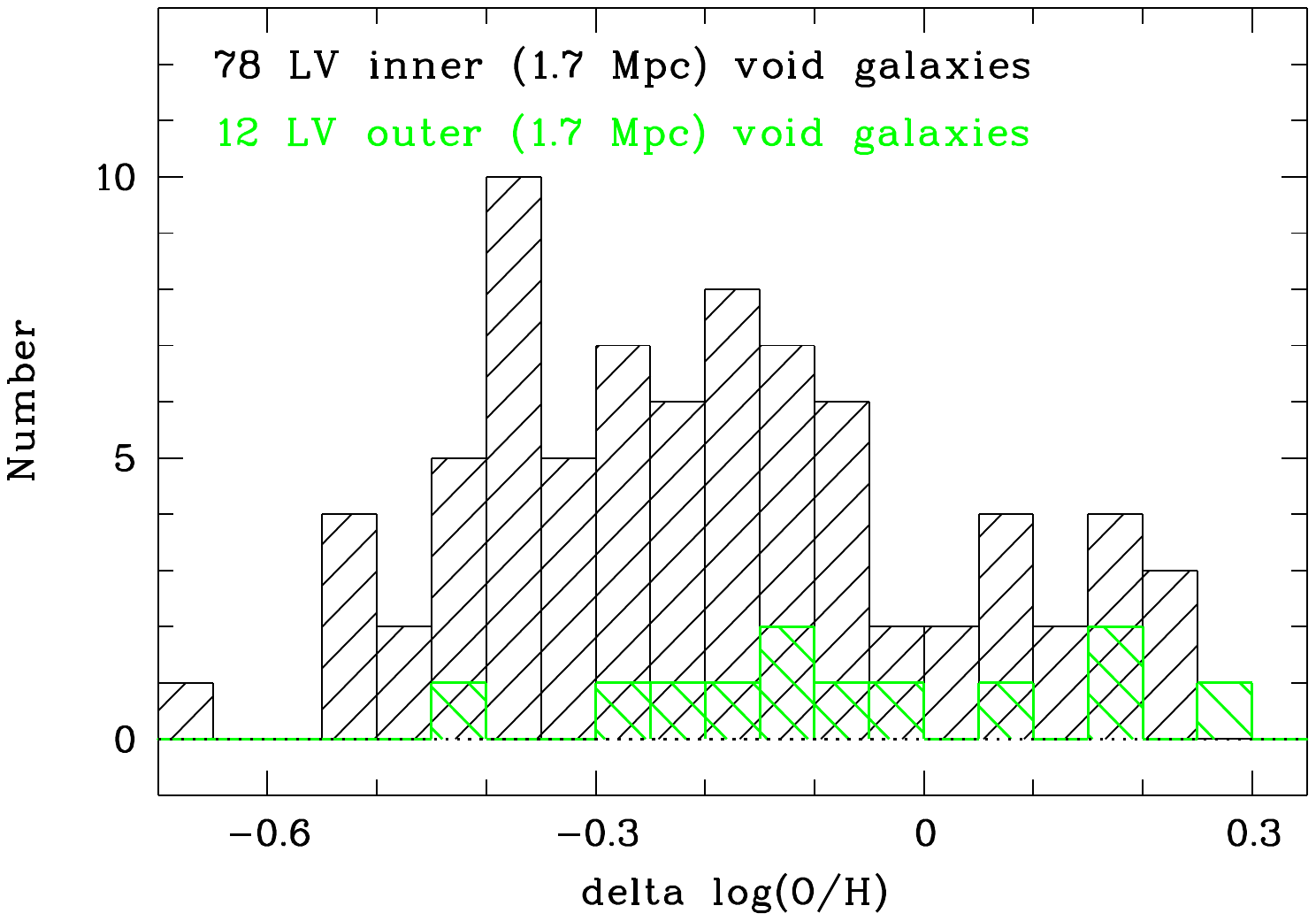}
\caption{The same data as in Figure~\ref{fig:ZvsMB_in+out}. The histograms show the differences of measured 12+log(O/H)
for the LV void galaxies and those expected from the reference relation from \citet{Berg12}.
{\bf left-hand panel:} separate histograms for 'inner' (71 galaxies, black) and 'outer' (19 galaxies, {\bf green})
subsamples for the case of the threshold D$_{\rm NN}$ = 2.0~Mpc.
{\bf right-hand panel:} same as the left-hand panel, but for the case of the threshold D$_{\rm NN}$ = 1.7~Mpc.
Black histogram is for 78 'inner', while the green one -- for 12 'outer' galaxies. See text for more detail.
}
\label{fig:hist2}
\end{figure*}

The effect of the reduced metallicity in void galaxies, already mentioned in several our papers,
and its elevated scatter for a given galaxy luminosity/mass,
can be due to the interplay between various
factors. In particular, the chemical evolution of void galaxies and their current metallicity
can depend on their local environment and clustering, and on the faster or slower accretion
of the unprocessed intergalactic gas. Since void substructure resembles that of
a mini-Universe \citep{Gottlober03,Aragon13}, void galaxies can be associated
with more massive hosts or to be well isolated. In general, the local environment is expected
to affect the secular evolution, especially of the smaller-mass companions. To uncover these factors,
it takes a more careful analysis of the studied sample.

On the other hand, since we study the LV void galaxy sample without any additional bias, we
can, in particular, examine, whether the outer and inner void galaxies show differences
in their properties. For this, we show outer and inner galaxies in the similar plot in
Figure~\ref{fig:ZvsMB_in+out}. While the statistics is still rather limited, there is a hint
that the 'outer' void objects show a smaller scatter in log(O/H) and follow closer the reference
relation of \citet{Berg12}. While the criterion of separation of outer void subsample by the inequality
of D$_{\rm NN} <$ 2~Mpc, is somewhat arbitrary, it seems to catch a transition of galaxy properties
from the most rarefied regions to a denser environment. Therefore, one can think that the large scatter
of void galaxies on log(O/H) in Figure~\ref{fig:ZvsMB_LV} can be partly explained by the contribution
of the outer objects.

In Figure~\ref{fig:hist2}, we show the distribution of differences in log(O/H) for 90 LV
void galaxies, separately for the 'inner' (black) and 'outer' (green) subsamples.
The left-hand panel shows the
histograms for the threshold  D$_{\rm NN}$ = 2.0~Mpc, as adopted in the original paper on the NVG sample.
The mean for the 71 'inner' galaxies is --0.193~dex with rms=0.202~dex. For the 19 'outer' galaxies, the mean
is --0.098~dex, with the rms=0.222~dex. While the scatter for both groups is similar, the 'inner' group has
a substantially more reduced O/H.

To check, how this effect depends on the adopted threshold for D$_{\rm NN}$, we divided these 90 LV void
galaxies into the 'inner' and 'outer' subsamples, taking the threshold D$_{\rm NN}$ = 1.7~Mpc. This means that
the selected 'outer' void galaxies are in average somewhat closer to the bordering large galaxies. The histograms
in the right-hand panel show the difference of log(O/H) for this case. For the 78 'inner' galaxies, the mean
is --0.191~dex with rms=0.204~dex, practically the same as for the previous histogram. For the 12 'outer'
galaxies, the mean is --0.055~dex, with rms=0.208~dex. That is, void galaxies with the smaller D$_{\rm NN}$
show the tendency to have the smaller difference of O/H with the reference relation, albeit the scatter is
larger than 0.15~dex characteristic of the reference sample of \citet{Berg12}. We notice that the statistics is still
limited, so the future data, incorporating the majority of the LV void galaxies, hopefully will result in more
firm conclusions.

\subsection{New dTr galaxies in void environment}
\label{sec:dTr}

Several studied void galaxies appear on their properties similar to the
handful known voids dwarfs of transitional type (dTr) \citep[][and references therein]{KK258, KK242}.
They show very little current SF as evidenced either by non-detected H$\alpha$ emission, or a single
region with the faint H$\alpha$. At the same time, their \HI\ content does not differ much from
dIr galaxies of the same luminosity. The unbiased study of the LV void dwarfs allows one to extend
the sample of these rare objects and to address the issue of their origin on the larger statistical data.

Below we enumerate the void galaxies, according to the new data, falling into this category with
the higher or lower confidence. Several of them are found as \HI-sources in the blind ALFALFA survey
\citep{ALFALFA}. A few similar ALFALFA galaxies are found in the SHIELD project described by
\citet{Cannon11,SHIELD,McQuinn14}.

{\bf PGC712531 = J033903.0-304921}. This bluish dwarf shows only a faint H$\alpha$ in emission.

{\bf AGC225197 = EVCC1184 = J124942.11+052922.} A blue low-surface brightness (LSB) dwarf
with the non-central blue compact object.
  The long slit was positioned to this blue object. Spectrum of this galaxy in SDSS DR12  \citep{DR12}
  looks similar to our.

{\bf AGC227972 - J125024.0+045422}. A bluish oval galaxy without prominent SF regions.

{\bf PGC2807150 = KKH86 = AGC231980.} This LSB dwarf is rather bright \HI\ source,
with F(\HI) (ALFALFA) = 0.8~Jy~\kms\ and the ratio of M(\HI)/L$_{\rm B} \sim$0.75.
At the same time, the on-line database of \citet{UNGC} provides its H$\alpha$-net image
with no tracers of emission. Our value of its V$_{\rm hel}$ = 298$\pm$40~\kms\
comes from the wavelengths of the Balmer absorption lines in the body.

\subsection[]{Issue of mistaken redshifts in the NVG sample}
\label{ssec:mistaken_NVG}

As noticed in the Introduction, the great majority of the NVG sample galaxies
in PTM19 are taken from the HyperLEDA data base.
On results of our spectral observations of a hundred NVG galaxies (that is residing
in the nearby voids), presented in \citet{XMP.SALT,XMP.BTA} and in this work,
we found that
$\sim$15~per cent of them have velocities different from the catalog (HyperLEDA)
values much larger than one expects from the cited uncertainties. The great majority
of them comes from objects in the Southern hemisphere. Of these, the majority of objects
with wrong velocities appear in the HyperLEDA from the 2dFGRS galaxies with redshifts
derived via absorption lines.

This allows us to formulate a caution for the use of the 2dFGRS redshift data for
statistical studies of galaxies, at least in the nearby Universe, where we
have checked many of their radial velocities.

\section[]{Summary and Conclusions}
\label{sec:conclusions}

In the previous papers \citep{XMP.SALT,XMP.BTA} we presented results of spectroscopy
for 46 objects from the sample of 60 preselected {\it candidate} XMP void objects
from the fainter part ($M_{\rm B} \gtrsim -14.3$) of the Nearby Void Galaxies (NVG) sample.
Ten of them were found to be the XMP objects, that is having 12+log(O/H) $\lesssim$ 7.21~dex.
13 more new void galaxies were found somewhat less metal-poor, with 12+log(O/H) $\lesssim$ 7.35~dex.

In this work we observed 8 galaxies of the remaining 13 XMP candidates and reobserved
one XMP dwarf from \citet{XMP.SALT}. Two new XMP dwarfs are found of these eight objects.
Two additional dwarf galaxies with the Oxygen abundance of 12+log(O/H) $\lesssim$ 7.35~dex are found as well.
Two known XMP void dwarfs were reobserved to improve the accuracy of their metallicity.

The rest 52 observed galaxies are selected to reside in the LV. They represent mostly a fainter part
of the total subsample of the LV void galaxies. The currently updated list comprises of 243 LV void objects.
Of 39 objects with the well detected strong lines, nine new void galaxies appeared to be very metal-poor,
with 12+log(O/H) $\lesssim$ 7.39~dex. Of them, two more XMP dwarfs are found, with 12+log(O/H) = 7.07
and 7.20~dex.

Summarising the results, presented here and the related discussion, we draw the
following conclusions:

\begin{enumerate}
\item
The spectroscopy of 8 remaining  'candidate' XMP dwarfs from the sample of 60
objects residing in the nearby voids, results in the discovery of two additional XMP dwarfs,
with 12+$\log$(O/H) $\sim$7.19 -- 7.20~dex ($\sim$Z\sunn/30).
\item
The almost completed program of search for void XMP dwarfs results in 11 new
objects with Z$_{\rm gas}$ = \mbox{Z\sunn/50 -- Z\sunn/30}, that comprises $\sim$20~per cent
of the selected candidates. In addition, 23 very low-metallicity dwarfs
(Z$_{\rm gas}$ = Z\sunn/30 -- Z\sunn/20) are found, that is $\sim$40~per cent of the initial list.
This finding increases substantially the number of the known nearby
very metal-poor dwarfs and allows us to conduct a deeper study of their
individual and group properties and of their possible relation to Very Young Galaxies.
\item
The first results are presented for the on-going project of the studying
a subsample of 243 NVG galaxies residing in the Local Volume. Of 52 observed objects only 37
appeared in real to reside in the nearby voids, while the rest 15 had the wrong radial velocities
in HyperLEDA.
\item
The strong lines of Oxygen are detected and its abundance is estimated in 32 of these
37 NVG objects, with the total range of 12+log(O/H) of 7.07~dex to $\sim$8.0~dex.
\item
The addition of the other available data on the LV void dwarfs allows us to probe
the relation between 12+log(O/H) and M$_{\rm B}$ on a sample of $\sim$90 objects.
The general trend of the reduced metallicity for a given luminosity, already
known from our earlier studies of the nearby void galaxies, is also well seen on the
LV void galaxies. Besides, on this compilation of the LV void galaxies, we find a large
scatter in O/H for a fixed M$_{\rm B}$.
Partly, the reason of this effect can be the presence in the LV void sample
of galaxies in the 'bordering' regions of voids.
\item
The studied void galaxies show a wide range of the star-forming activity.
While the most typical dwarfs, dIrrs and late Spirals, show several SF regions,
we find four new dTr galaxies with the substantial neutral gas reservoir
and the absent or very subtle signs of the current/recent SF.
\item
Two Local Volume \HI\ objects, HIPASS~J1614--72 and AGC208329 appear to have
a mistaken optical identification. The search for their alternative optical
counterparts did not reveal potential candidates. Since the full census of the
LV population has the important implications for comparison of the LV with cosmological
simulations, the understanding of the nature of these \HI\ objects will take special
efforts, including their \HI-mapping and the improvement of their position accuracy.
\end{enumerate}

\section*{Acknowledgements}
This work is based on observations obtained with the Southern African Large
Telescope (SALT), programs \mbox{2017-2-MLT-001}, \mbox{2020-2-MLT-005},
\mbox{2022-1-MLT-003}  (PI: Kniazev). The reported study was funded by Russian
Science Foundation according to the research project  22-22-00654.
AYK acknowledges support from the National Research Foundation (NRF) of
South  Africa. We thank the anonymous reviewer for the useful comments and
suggestions, which allowed us to improve the content and clarity of the paper.
The use of the HyperLEDA database % \footnote{http://leda.univ-lyon1.fr}
is greatly acknowledged.
This research has made use of the NASA/IPAC Extragalactic Database (NED)
which is operated by the Jet Propulsion Laboratory, California Institute
of Technology, under contract with the National Aeronautics and Space
Administration. We also acknowledge the great effort of the ALFALFA team
which opened access to the nearby Universe gas-rich dwarfs with low or
moderate SFR and thus helped us to identify the majority of very low
metallicity galaxies of this study.

We also acknowledge the use of the SDSS and Legacy surveys databases.
Funding for the Sloan Digital Sky Survey (SDSS) has been provided by the
Alfred P. Sloan Foundation, the Participating Institutions, the National
Aeronautics and Space Administration, the National Science Foundation,
the U.S. Department of Energy, the Japanese Monbukagakusho, and the Max
Planck Society. The SDSS Web site is http://www.sdss.org/.
The SDSS is managed by the Astrophysical Research Consortium (ARC) for the
Participating Institutions.

The DESI Legacy Imaging Surveys consist of three individual and complementary projects:
the Dark Energy Camera Legacy Survey (DECaLS), the Beijing-Arizona Sky Survey (BASS), and
the Mayall z-band Legacy Survey (MzLS). DECaLS, BASS and MzLS together include data obtained,
respectively, at the Blanco telescope, Cerro Tololo Inter-American Observatory, NSF's NOIRLab;
the Bok telescope, Steward Observatory, University of Arizona; and the Mayall telescope, Kitt
Peak National Observatory, NOIRLab.
NOIRLab is operated by the Association of Universities for Research in Astronomy (AURA) under
a cooperative agreement with the National Science Foundation. Pipeline processing and analyses
 of the data were supported by NOIRLab and the Lawrence Berkeley National Laboratory (LBNL).
Legacy Surveys also uses data products from the Near-Earth Object Wide-field Infrared Survey
Explorer (NEOWISE), a project of the Jet Propulsion Laboratory/California Institute of Technology,
funded by the National Aeronautics and Space Administration. Legacy Surveys was supported by:
the Director, Office of Science, Office of High Energy Physics of the U.S. Department of Energy;
the National Energy Research Scientific Computing Center, a DOE Office of Science User Facility;
the U.S. National Science Foundation, Division of Astronomical Sciences; the National Astronomical
Observatories of China, the Chinese Academy of Sciences and the Chinese National Natural Science
Foundation. LBNL is managed by the Regents of the University of California under contract to the
U.S. Department of Energy.

\section*{Data Availability}
The data underlying this article are available in Appendices~A, B and C, which are
available only in the on-line supplementary materials  of the paper.
%===========================================================================

\label{lastpage}

\end{document}

%% file: tab_jour1.tex
\setcounter{qub}{0}
\begin{table} % [htbp]
\begin{center}
\caption{Journal of SALT spectral observations. Main sample}
\label{tab:journal1}
\hoffset=-3.5cm
\begin{tabular}{r|l|l|l|r|c|c} \hline % \\ % [-0.2cm]
\MC{1}{r}{No} &
\MC{1}{c|}{Name} &
\MC{1}{c|}{Date} &
%\MC{1}{c|}{Grating} &
\MC{1}{c|}{Expos.}&
\MC{1}{c|}{PA} &
%\MC{1}{c|}{Off-set star}&
\MC{1}{c|}{$\theta$\arcsec}&
\MC{1}{c}{Air}  \\

\MC{1}{r|}{} &
\MC{1}{c|}{ } &
\MC{1}{c|}{ } &
%\MC{1}{c|}{ } &
\MC{1}{c|}{time, s}&
\MC{1}{c|}{ } &
%\MC{1}{c|}{J2000 coord.} &
\MC{1}{c|}{} &
\MC{1}{c}{mass}\\

%\MC{1}{r|}{} &
%\MC{1}{c|}{1} & \MC{1}{c|}{2} & \MC{1}{c|}{3} & \MC{1}{c|}{4} &
%\MC{1}{c|}{5} & \MC{1}{c|}{6} & \MC{1}{c}{7} \\
%\\[-0.2cm]
\hline \\[-0.2cm]
\qq&J0015+0104         & 2019.10.01  &2$\times$1200& 346.0 & 1.6 & 1.22 \\ %                            & 00:15:??.??+01:04:??.?
\qq&ESO294-010$^+$     & 2020.12.18  &2$\times$1200&  64.0 & 1.4 & 1.21 \\ % J002633.4-415120 LV O/H=7.69--7.82+-0.18 (se) depend. on CHb_EWabs after ULySS fit
\qq&PGC004055$^+$      & 2022.06.27  &2$\times$1200&  66.0 & 1.7 & 1.27 \\ % J010822.0-381233 LV O/H~7.30+-.05 (s,c aver.2 knots)       =KK011  Vh=654
\qq&UGC01085$^+$       & 2022.11.22  &2$\times$1100& 169.0 & 1.6 & 1.32 \\ % J013118.9+074716 LV O/H=7.77+-.11 (mse) MB = -13.44 Vd=727
\qq&AGC124137$^*$      & 2018.11.10  &2$\times$1200& 238.0 & 1.6 & 1.34 \\ %                            & 02:31:37.0+09:31:44
   &-\#-               & 2019.09.02  &2$\times$1200& 238.0 & 1.5 & 1.35 \\ %                            & 02:31:37.0+09:31:44
   &-\#-               & 2019.10.25  &2$\times$1200& 238.0 & 1.7 & 1.34 \\ %                            & 02:31:37.0+09:31:44
\qq&ESO199-007$^+$     & 2021.02.03  &2$\times$1200&  13.0 & 1.5 & 1.33 \\ % J025804.1-492256 LV O/H=7.31+-.05 (s,c bri) ~7.18 (s,c fai) M_B~-12.5
\qq&PGC1166738$^*+$    & 2019.09.06  &2$\times$1150&  41.5 & 1.6 & 1.22 \\ % J030646.9+002811 LV O/H=7.10+-.07 (s) b-knot; 7.28+-.06 (s) a-knot (sum on 4 spectra)
%  &-\#-               & 2019.12.18  &1$\times$1150&  ??.0 & 1.? & 1.29 \\ %                                 &
   &-\#-               & 2019.12.25  &2$\times$1100&  41.5 & 1.6 & 1.29 \\ %                                 &
   &-\#-               & 2019.12.29  &2$\times$1100&  41.5 & 1.7 & 1.29 \\ %                                 &
\qq&PGC013294$^*+$     & 2019.09.01  &2$\times$1200& --24.0& 1.1 & 1.27 \\ % J033556.8-451129 LV+S2 O/H=7.65+-0.0? (Te, all lines!) 2 HII regions from HST image. PA=-24 D=7.3
\qq&PGC712531$^+$      & 2021.07.31  &2$\times$1200& 101.5 & 2.0 & 1.22 \\ % J033903.0-304921 LV? PG2300 2dF V=839+-123 Only possib. line Ha at V=615 low S/N!
\qq&PGC681755$^+$      & 2022.12.26  &2$\times$1300&  51.0 & 1.2 & 1.27 \\ % J033955.0-330309 LV PG900 2*1200s BCG! E+A spect. O/H_mse,c=7.71+-0.12. =FCCB1379 V(HI)=745+-21 V_emis = 720+-17
\qq&ESO359-024$^+$     & 2023.01.24  &2$\times$1200& 255.5 & 1.7 & 1.28 \\ % J041057.5-354951 LV PG900 2*1200s Sm SIGRID19 M_B=-14.8  O/H_d=7.44+-0.13
\qq&PGC016389$^+$      & 2023.02.22  &2$\times$1200&  92.0 & 1.0 & 1.19 \\ % J045701.2-424803 LV PG900 2*1200s
\qq&HIJ0517-32$\dagger$$^+$& 2022.01.06  &2$\times$1200& 345.0 & 2.3 & 1.31 \\ % J051721.6-324535 LV pg900 star-burst Vh=798+-6, 4363! O/H=7.93+-.05 (Te,aver) MB~-14.27 V=818,821+-6 (BUT need bary correct)
\qq&ESO553-046$^+$     & 2023.01.25  &2$\times$1250& 101.0 & 2.0 & 1.27 \\ % J052705.8-204040 LV pg900 SIGRID32 M_B=-14.71 D=6.70  O/H_d = 8.02+-0.05
\qq&PGC138836$^+$      & 2021.02.04  &2$\times$1150& 161.0 & 2.0 & 1.31 \\ % J055735.2+072913 LV O/H=7.53+-0.07 (mse) pg900 M_B_corr =-13.82!
   &-\#-               & 2021.02.07  &2$\times$1150& 161.0 & 2.0 & 1.30 \\ %                     O/H=7.54+-0.06 (mse) pg900
\qq&PGC018431$^+$      & 2023.04.11  &2$\times$1200&  87.0 & 1.8 & 1.23 \\ % J060719.7-341215 LV O/H=7.70+-0.06 Te    pg900
\qq&ESO308-022$^+$     & 2023.02.21  &2$\times$1200& 331.0 & 1.9 & 1.28 \\ % J063932.9-404317 LV O/H~7.3-7.4  pg900, 2*1200s
\qq&PGC020125$^+$      & 2022.12.18  &2$\times$1200&  70.0 & 1.1 & 1.18 \\ % J070517.7-583108 LV O/H_s,c=7.18 & 7.38 (2 knots) Argo dIrr =11HUGS115 D=5.3 Vh=564 M_B=-14.16
   &-\#-               & 2022.12.26  &2$\times$1200&  70.0 & 1.2 & 1.21 \\ % addit.spectrum to improve S/N O/H_s,c = 7.21 & 7.35
\qq&J0935-1348         & 2013.12.28  &3$\times$800 &   0.0 & 1.7 & 1.25 \\ % J093521.6-134852 O/H=7.28+-0.07 (s) LONG AGO! on Monoceros void program, DNN=1.94+-0.60                                  & 10:01:03.95+08:47:48
\qq&PGC029033$^+$      & 2023.01.21  &2$\times$1100&  33.0 & 1.8 & 1.23 \\ % J100138.21-081455 LV PG900 faint Ha,5007 & very faint Hb (em+abs)
\qq&PGC041667$^+$      & 2022.04.27  &2$\times$1100& 293.5 & 1.5 & 1.27 \\ % J123307.9-003158  LV PG900 O/H_mse,c ~7.8
\qq&PGC091215$^+$      & 2022.03.05  &2$\times$1300&   5.0 & 1.3 & 1.23 \\ % J123655.0+013654  LV Edge-on, 3 EL regions, poor? 7.35 or 7.75? M_B=-12.65
\qq&AGC227970$^+$      & 2022.05.03  &2$\times$1200&  40.0 & 1.9 & 1.30 \\ % J124601.4+042252  LV PG900 comet V=643 O/H_dir = 7.71+-.08  O/H_mse,c=7.79+-.12
\qq&AGC225197$^+$      & 2022.03.05  &2$\times$1200& 352.0 & 1.3 & 1.27 \\ % J124942.1+052922  LV? J1249+0529 gas-rich only Balm. abs M_B=-11.77 Vh(HI)=738+-6 (V_SDSS+abs = 680?) Vh(abs.SALT)=866
\qq&UGC07983$^+$       & 2022.04.27  &2$\times$1100&   5.0 & 1.1 & 1.33 \\ % J124947.0+035032  LV PG900 O/H_mse,c ~7.76 (aver.)
\qq&PGC1289726$^+$     & 2022.04.25  &2$\times$1100&  38.0 & 1.3 & 1.27 \\ % J124959.2+054915  LV PG900 prelim. O/H_mse,c=7.91+-0.09
%\qq&PGC135809          & 2021.04.04  &2$\times$1200&  ??.0 & 1.? & 1.?? \\ % J125004.8-001357  LV Ha_image (faint Ha knot on W edge)
%   &-\#-               & 2022.05.27  &2$\times$1150&  ??.0 & 1.? & 1.?? \\ % Ha_ima (2nd time?) 1"-seeing
%   &-\#-               & 2023.01.22  &2$\times$1100&  ??.0 & 1.? & 1.?? \\ % PG900 v.faint cont.light. slit position was wrong Vh=754 (Arecibo)
%\qq&AGC227972          & 2021.06.12  &2$\times$1200& 134.5 & 1.? & 1.?? \\ % J125024.0+045422  LV VLSB Ha_image; 2 knots VHI=650
\qq&AGC227972$^+$      & 2022.04.26  &2$\times$1100&  134.5 & 1.9 & 1.29 \\ % PG900 2 faint Ha-knots @PA~-20? V1=481 V2=670 dV=190 km/s dY12~6" => 0.27 kpc
\qq&AGC227973$^+$      & 2021.04.07  &2$\times$1100&  326.0 & 1.4 & 1.28 \\ % J125039.9+052052  LV O/H_s=7.00-7.07 (PG900 2 spectra with shifted slit)
   &-\#-               & 2021.06.13  &2$\times$1100&  326.0 & 1.6 & 1.26 \\ % O/H_s=7.16?     (repeat PG900)
   &-\#-               & 2022.03.09  &2$\times$1100&  326.0 & 1.3 & 1.28 \\ % O/H_s=7.08 PG900
%\qq&AGC226122          & 2022.05.30  &2$\times$1100&  ??.0 & 1.? & 1.?? \\ % J125215.4+042727  LV Ha-ima LSBD, faint Ha-arc? near center, + knot @W edge?
\qq&AGC226122$^+$      & 2023.01.31  &2$\times$1100&   46.5 & 1.6 & 1.29 \\ % PG900 faint spectrum!
\qq&PGC1264260$^+$     & 2021.02.06  &2$\times$1100&   23.0 & 1.6 & 1.24 \\ % J125343.3+040914  LV? PG0900 D(Vd)~10 Mpc MB~-12.7 SDSS spec. no 3727, O/H=7.56+-.08 (dir)
\qq&UGC08055$^+$       & 2022.04.30  &2$\times$1100&   97.0 & 1.3 & 1.33 \\ % J125604.4+034846  LV  PG900 O/H_dir = 7.71+-.32, O/H_mse,c=7.79+-.12
\qq&PGC044681$^*+$     & 2023.02.23  &2$\times$1100&   30.5 & 1.8 & 1.30 \\ % J125956.6-192441  LV  PG900 PA=30.5? slit thru AGN(??) & blue NE knot (Legacy chart)
\qq&KKH86$^+$          & 2022.05.31  &2$\times$1100&  324.5 & 1.3 & 1.26 \\ % J135433.4+041440  pg900 Only Bal. absorbts
%   &-\#-               & 2022.05.27  &2$\times$1100&  ??.0 & 1.? & 1.?? \\ % Ha image
\qq&AGC716018$^*$      & 2020.03.16  &2$\times$1150&   41.0 & 1.6 & 1.30 \\ % J1430+0709 S2 O/H~7.77 (se,c)
\qq&AGC249197$^*$      & 2019.05.29  &2$\times$1150&  190.0 & 1.0 & 1.36 \\ % J1449+0956 S2 O/H=7.78+-0.08 (se,c) model contin. subtract PG0900 ~2300s, or altern. O/H=7.40 (s,c)! much depends on IHb correct. for EW(abs)
\qq&HIJ1738-57$\dagger$$^*+$& 2019.05.03  &2$\times$1150&   84.5 & 1.2 & 1.23 \\ % J173842.9-571525 LV+S2 Bri: O/H=7.61 (se,c)+-0.07 Fai: O/H=7.46 (se,c)+-0.07 or 7.35+-0.07 (s,c)
\qq&PGC408791$^+$      & 2021.07.29  &2$\times$1150&  279.5 & 1.4 & 1.27 \\ % J202608.8-552950 LV PG2300 V(6dF)=893+-67
  &-\#-                & 2021.11.06  &2$\times$1150&  279.5 & 2.0 & 1.27 \\ % PG900, V=836+-10 => Vd=500; O/H(E-knot)_s ~7.40+-0.05 dex (or a bit smaller)
\qq&PGC129680$^+$      & 2022.04.28  &2$\times$1100&   45.5 & 1.0 & 1.29 \\ % J210804.8-471942 LV  PG900 O/H_mse~7.99+-0.1
% &-\#-                & 2022.07.20  &2$\times$1100&   ??.0 & 1.? & 1.?? \\ % PG900 (by error addit. observations)
\qq&PGC1016598$^*$     & 2019.07.11  &2$\times$1300&  103.0 & 1.8 & 1.30 \\ % PG900 J213902.9-073443 KnotB(W): O/H=7.30+-.07 (s) KnotA(E): O/H=7.58+-.07 strong overlap. +SDSS spectrum
\qq&ESO289-020$^+$     & 2022.04.28  &2$\times$1100&  331.5 & 1.4 & 1.23 \\ % J222111.7-454035 LV PG900 O/H_mse,c = 7.73 (aver. 2 knots)
% &-\#-                & 2022.07.25  &2$\times$1100&  ??.0 & 1.? & 1.?? \\ %
\qq&ESO238-005$^*+$    & 2019.05.11  &2$\times$1100&  47.0 & 1.6 & 1.22 \\ % J222230.5-482414 S2 O/H=7.32+-0.06(s)  N/O~-1.20 D(TRGB)=8.0
\qq&PGC1028063$^+$     & 2022.05.26  &2$\times$1100& 300.0 & 1.2 & 1.20 \\ % J224223.4-065010 LV PG900 O/H_Te = 7.79+-0.09
\qq&ESO347-017$^+$     & 2022.05.27  &2$\times$1100&  90.0 & 1.2 & 1.21 \\ % J232656.2-372049 LV O/H=7.82+-0.03 Te  4363! O/H - high. 4959/Hb~1.3
\qq&PGC3192333$^+$     & 2020.12.17  &2$\times$1100& 265.5 & 1.4 & 1.22 \\ % J234133.8-353023 LV O/H=7.26+-0.08 (Te) or 7.16 (s,c) check Vh  orig.dV=120+-125
\qq&PGC680341$^+$      & 2020.12.15  &2$\times$1100&  51.5 & 1.2 & 1.24 \\ % J234147.5-330841 LV O/H=7.34+-0.04 (Te) check Vh
\qq&ESO348-009$^+$     & 2022.05.31  &2$\times$1200&  90.0 & 1.3 & 1.31 \\ % J234923.4-374622 LV pg900, O/H_s=7.45 ???
\qq&ESO149-003$^+$     & 2022.06.06  &2$\times$1200& 146.5 & 1.6 & 1.30 \\ % J235202.8-523438 LV pg900 O/H_Te=7.70+-0.04   M_B=-14.2
\hline
%\hline \hline \\[-0.2cm]
\multicolumn{7}{p{8.5cm}}{$^+$ LV void galaxies; $^*$ XMP candidates from PEPK20; } \\ %
\multicolumn{7}{p{8.5cm}}{$\dagger$ 'HIJ' means 'HIPASSJ'} \\ %
\end{tabular}
\end{center}
\end{table}

%% file: tab_jour2.tex
\setcounter{qub}{0}
\begin{table}
\begin{center}
\caption{Journal of SALT spectral observations. Objects with wrong velocities}
\label{tab:journal2}
\hoffset=-2cm
\begin{tabular}{r|l|l|l|r|c|c} \hline  \hline \\ [-0.2cm]
\MC{1}{r|}{No.} &
\MC{1}{c|}{Name} &
\MC{1}{c|}{Date} &
\MC{1}{c|}{Expos.}&
\MC{1}{c|}{PA} &
\MC{1}{c|}{$\theta$\arcsec}&
\MC{1}{c}{Air}  \\

\MC{1}{r|}{} &
\MC{1}{c|}{ } &
\MC{1}{c|}{ } &
%\MC{1}{c|}{ } &
\MC{1}{c|}{time, s}&
\MC{1}{c|}{ } &
%\MC{1}{c|}{J2000 coord.} &
\MC{1}{c|}{} &
\MC{1}{c}{mass}\\

%\MC{1}{r|}{} &
%\MC{1}{c|}{1} & \MC{1}{c|}{2} & \MC{1}{c|}{3} & \MC{1}{c|}{4} &
%\MC{1}{c|}{5} & \MC{1}{c|}{6} & \MC{1}{c}{7} \\
\\[-0.2cm] \hline \\[-0.2cm]
\qq&PGC3197756$^+$     & 2021.09.10  &2$\times$1200&  295.0 & 1.8 & 1.27 \\ % J001736.3-312618 no-LV PG2300 2dF Vemi=749+-123 V(Ha,SII) ~2700  &
\qq&PGC901638$^+$      & 2021.08.03  &2$\times$1200&  314.5 & 1.3 & 1.23 \\ % J002951.4-160954 no-LV PG900 2dF Vabs=724+-59  z~0.092 OII3727 & abs. (CaH,K; Hg,Hb, MgI5180)
\qq&PGC3207684$^+$      & 2021.10.08  &2$\times$1150&  259.0 & 1.3 & 1.22 \\ % J020140.1-291749 no-LV PG900 2dF Vabs=660, z=0.109
\qq&PGC3210819$^+$      & 2021.08.04  &2$\times$1200&  317.0 & 1.8 & 1.22 \\ % J022308.8-295233 no-LV 2dF V=779 z=0.16274
\qq&PGC504827$^+$      & 2021.08.01  &2$\times$1200&  282.5 & 1.3 & 1.25 \\ % J024024.4-471257 no-LV 2dF V=812+-59  Ha @V~1800 km/s
%\qq&PGC645417          & 2021.08.04  &2$\times$1150&   48.0 & 1.? & 1.28 \\ % J032733.9-354303 LV? PG2300 V_ori(Ha, EW=0.3)=723+-79   poor spectrum nothing! & 02:47:09.46+10:06:46
\qq&PGC645417$^+$       & 2022.01.23  &2$\times$1150&   48.0 & 2.2 & 1.28 \\ % PG2300 Faint Ha at z=0.0333                               &
\qq&PGC016383          & 2023.02.22  &2$\times$1200&   92.0 & 1.0 & 1.19 \\ % J045701.2--424803 NON-VOID
\qq&AGC208329$^+$      & 2019.01.06  &2$\times$1200&   94.5 & 1.5 & 1.29 \\ % J1015+0335 NON-VOID O/H=8.18+-0.18 Te, or O/H=7.80+-0.08 (se) PG900 4363 Vopt~1335 vs VHI~1018 near lum.bording galaxy!
%\qq&AGC225197          & 2022.03.05  &2$\times$1200&  ??.0 & 1.? & 1.?? \\ % J124942.1+052922 LV? gas-rich only Balm. abs M_B=-11.77 Vh(HI)=738+-6 (V_SDSS_abs~680?) Vh(abs.SALT)=866
\qq&J1355+04B$\dagger$ & 2022.05.31  &2$\times$1100&   73.0 & 1.0 & 1.31 \\ % J1354??.?+0412?? no-LV pg900 Vh~5100, MB~-14, O/H~7.5??
\qq&PGC1069207$^+$     & 2021.04.14  &2$\times$1100&  137.0 & 1.2 & 1.21 \\ % J143011.3-033552 no-LV PG2300 V(Ha)=988+-9 vs V(2dF)=690+-123 => Vd=892 D=12.2+-?? MB_new=-12.5
\qq&PGC264615$^+$      & 2021.04.04  &2$\times$1200&   21.0 & 1.3 & 1.32 \\ % J161354.9-721446 no-LV wrong opt.counterpart VHI~300; z=0.0695 MB~-19.3!
\  &-\#-               & 2022.04.12  &2$\times$1200&   21.0 & 1.5 & 1.35 \\ % PG1800 LINER? shock excit.: strong NII,SII,OI6300
\qq&PGC064718$^+$      & 2023.04.02  &2$\times$1200&   52.0 & 1.4 & 1.31 \\ % J202733.8-550525 no-LV V(LEDA)~850, V(SALT) ~10900
\qq&PGC162688$^+$      & 2021.08.02  &2$\times$1200&  107.0 & 1.6 & 1.26 \\ % J204847.3-121654 no-LV PG900 S0-a 6dF V=843+-45 z~0.01956 => V~5900   O/H~7.50?
\qq&PGC163318$^+$      & 2021.07.17  &2$\times$1200&   43.0 & 2.0 & 1.23 \\ % J213935.2-401653 no-LV V(2dF)=1034+-59 => Vd=762  PG2300: absorb. at z~0 *-like
\qq&6dFJ2226-2916$^+$  & 2020.11.13  &2$\times$1200&   23.0 & 1.8 & 1.21 \\ % J2226-2916       no-LV Distant Sey1.5? (z~0.35) confusion of broad Ha and Hb/N1+N2
\qq&ABELL3888.14$^+$   & 2021.07.30  &2$\times$1200&  337.5 & 1.3 & 1.24 \\ % J223329.3-372731 no-LV V=1008+-19  our V_abs=-90
\qq&ABELL3888.12$^+$   & 2021.08.03  &2$\times$1200&  134.0 & 1.5 & 1.23 \\ % J223633.3-373005 no-LV PG900 V=760+-13 Vabs *-like z=0.388 (OII3727, Hd,Hg,CaH,K etc)
\hline \hline \\[-0.2cm]
\multicolumn{7}{p{8.0cm}}{$^+$ selected as the LV void galaxies} \\
\multicolumn{7}{p{8.0cm}}{$\dagger$ \citet{MU} adopt it as a companion of KKH86} \\
\end{tabular}
\end{center}
\end{table}

%% file: tab_mistake.tex
\setcounter{qub}{0}
\begin{table*}
\begin{center}
\caption{SALT results for objects with wrong velocities}
\label{tab:wrong}
\hoffset=-2cm
\begin{tabular}{r|l|l|r|r|c|c|l} \hline  \hline \\ [-0.2cm]
\MC{1}{r|}{No.} &
\MC{1}{c|}{Name} &
\MC{1}{c|}{J2000 coord.} &
\MC{1}{c|}{V$_{orig}$} &
\MC{1}{c|}{V$_{SALT}$}&
\MC{1}{c|}{B-mag} &
\MC{1}{c|}{M$_{\rm B}$}&
\MC{1}{c}{Notes}  \\

\MC{1}{r|}{} &
\MC{1}{c|}{ } &
\MC{1}{c|}{ } &
\MC{1}{c|}{\kms } &
\MC{1}{c|}{\kms}&
\MC{1}{c|}{ } &
\MC{1}{c|}{} &
\MC{1}{c|}{}    \\

\MC{1}{r|}{1} &
\MC{1}{c|}{2 } &
\MC{1}{c|}{3 } &
\MC{1}{c|}{4 } &
\MC{1}{c|}{5 }&
\MC{1}{c|}{6 } &
\MC{1}{c|}{7} &
\MC{1}{c|}{8 }    \\

\\[-0.2cm] \hline \\[-0.2cm]
%\hline \hline
\qq&PGC3197756\dg     & J001736.3--312618   &749$\pm$123& 2703\p10   &    19.28 &--13.64   & doubtful, too narrow H$\alpha$  \\ %  no-LV PG2300 2dF Vemi=749+-123 V(Ha,SII) ~2700, but W~1.3 A may be too narrow?
\qq&PGC901638         & J002951.4--160954   & 724$\pm$59& z = 0.092  &    17.74 &--20.30   & abs. spectrum \\ %  no-LV PG900 2dF Vabs=724+-59  z~0.092 OII3727 & abs. (CaH,K; Hg,Hb, MgI5180)
\qq&PGC3207684        & J020140.1--291749   & 660\p123  & z = 0.109  &    19.45 &--18.90   &               \\ % -10.10 Vd=575
\qq&PGC3210819        & J022308.8--295233   & 779\p89   & z = 0.16274&    18.85 &--20.30   &               \\ % -11.07 Vd=704
\qq&PGC504827\dg      & J024024.4--471257   & 812$\pm$29& 1828$\pm$15&    19.24 &--12.68   & only faint H$\alpha$  \\ %  no-LV 2dF V=812+-59  Ha @V~1800 km/s
\qq&PGC645417\dg      & J032733.9--354303   & 723$\pm$79&  9942\p20  &    16.84 &--18.86   & only H$\alpha$ \\ %  Ha in the second spectrum at V~9900
\qq&PGC016383         & J045701.2--424803   & 3443\p61  & 2953\p10   &    15.48 &--17.30   & on slit with PGC016389  \\ %
\qq&AGC208329         & J101531.9+033508    &1018$\pm$04& 1370$\pm$10&    19.42 &--12.1:   & 12+log(O/H)(mse)=7.82. see text \\ %  NON-VOID O/H_dir =8.18+-0.18, PG900 4363 Vopt~1370 vs VHI~1018 near lum.bording galaxy! V(HI)=540!
%\qq&AGC225197        & J124942.1+052922    &738$\pm$06 & 866$\pm$20 &    1?.?? &--1?.?    & abs. spectrum \\ %  non-LV? J1249+0529 gas-rich only Balm. abs M_B=-11.77 Vh(HI,ALFA)=739+-18, W_50=130+-37! (SDSS: abs.680+-35) Vh(abs.SALT)=866 => Vd=864, D=11.8. MB=~12.1 (in Vir-Cl)
\qq&[MU2012]J1355+04B & J135429.5+041237    & ...       & 4831\p9    &    19.5: &--14.6:   & 12+log(O/H)(mse)=7.56. see text  \\ %  no-LV pg900 Vh~4800, MB~-14, O/H~7.56: exp.~V$\sim$300,
\qq&PGC1069207\dg     & J143011.3--033552   &690$\pm$123& 988$\pm$9  &    18.34 &--12.50   & New D = 12.2 Mpc \\ %  no-LV PG2300 V(Ha)=988+-9 vs V(2dF)=690+-123 => Vd=892 D=12.2+-?? MB_new=-12.5
\qq&PGC264615         & J161354.9--721446   & 383\p11   & z = 0.0695 &    17.94 &--19.3    & LINER?  \\ %  no-LV wrong opt.counterpart VHI~300; z=0.0695 MB~-19.3! shock excit.: strong NII,SII,OI6300
\qq&PGC064718         & J202733.8--550525   &831$\pm$52 &10710$\pm$35&    15.24 &--21.2    & ......  \\ %  no-LV V(LEDA)~830, V(SALT) ~10710+-35
\qq&PGC162688         & J204847.3--121654   &843$\pm$45 &5860$\pm$12 &    17.12 &--17.6:   & ......  \\ %  no-LV PG900 S0-a 6dF V=843+-45 z~0.01956 => V~5900   O/H~7.50?
\qq&PGC163318\dg      & J213935.2--401653   &1034$\pm$59& z = 0.0645 &    17.25 &--19.95   & with close E-gal at z = 0.0630  \\ %  no-LV V(2dF)=1034+-59 => Vd=762  PG2300: absorb. at z~0 *-like  (M or carbon star)
\qq&6dFJ2226336-291728& J222633.6--291728   &1025$\pm$45& z = 0.315  &    19.57 &--21.0:   & AGN Sy1.5    \\ %  no-LV Distant Sey1.5? (z~0.35) confusion of broad Ha and Hb/N1+N2
\qq&ABELL3888$\_$14:[PSE2006]1816 & J223329.3--372731   &1008$\pm$19&--166\p10   &    18.49 & ...      & star-like \\ %  no-LV V=1008+-19  our V_abs=-90
\qq&ABELL3888$\_$12:[PSE2006]2246 & J223633.3--373005   & 760$\pm$13& z = 0.388  &    18.80 &--22.3:   & star-like \\ %  no-LV PG900 V=760+-13 Vabs *-like z=0.388 (OII3727, Hd,Hg,CaH,K etc)
\hline \hline \\[-0.2cm]
\multicolumn{7}{p{6.5cm}}{$\dagger$ grism PG2300, range $\sim$6060--6880~\AA} \\
\end{tabular}
\end{center}
\end{table*}